\documentclass{aa}

\usepackage{natbib}
\bibpunct{(}{)}{;}{a}{}{,}
\usepackage{graphicx}
\usepackage{epsfig}
\usepackage{amsmath,amssymb}
\usepackage[dvipdfm,breaklinks=true,bookmarks=true,hypertexnames=false,bookmarksnumbered=true,bookmarksopen=false,bookmarks=true,bookmarksnumbered=true]{hyperref}

\newcommand{\msun}{M_{\odot}}
\newcommand{\kms}{\, {\rm km\, s}^{-1}}
\newcommand{\h}{\,h_{\rm 70}}
\newcommand{\hmm}{\h^{-1}}
\newcommand{\hmmsun}{\hmm\msun}
\newcommand{\kpc}{\, {\rm kpc}}
\newcommand{\hmkpc}{\hmm\kpc}
\newcommand{\hmlsun}{\h^{-2}\, L_\odot}
\newcommand{\Mpc}{\, {\rm Mpc}}
\newcommand{\hmMpc}{\hmm\Mpc}
\newcommand{\der}{{\rm d}}
\newcommand{\rede}{{\mathcal E}}
\newcommand{\reffig}[1]{Fig.~\ref{#1}}

\newcommand{\mslsun}{\h\,(M/L)_\odot}
\newcommand{\vhalf}{\sigma_{\rm h_4}}
\newcommand{\slos}{\sigma_{\rm los}}
\newcommand{\mypm}[2]{^{+#1}_{-#2}}
\newcommand{\chired}{\chi^2/{\rm dof}}
\newcommand{\scrit}{\Sigma_{\rm crit}}

\newcommand{\gen}{{\it ``gen''}~}

\begin{document}

\title{Projection effects in cluster mass estimates : the case of MS2137-23}
\author{R. Gavazzi \inst{1,2}}
 \titlerunning{The triaxiality of MS2137-23}

\offprints{R. Gavazzi, \email{rgavazzi@ast.obs-mip.fr}}
\date{}
\institute{Institut d'Astrophysique de Paris, UMR 7095, 98bis Bd Arago,
  F-75014 Paris, France \and Laboratoire d'Astrophysique, OMP, UMR
  5572, 14 Av Edouard Belin, F-31400 Toulouse, France}  

\abstract{
  We revisit the mass properties of the lensing cluster of galaxies
  MS2137-23 and assess the mutual agreement between cluster mass
  estimates based on strong/weak lensing, X-rays and stellar dynamics.
  We perform a thorough elliptical lens modelling using arcs and their
  counter-images in the range $20\lesssim R\lesssim100$ kpc and weak
  lensing ($100\lesssim R\lesssim1000$ kpc). We confirm that the dark
  matter distribution is well consistent with an NFW profile with high
  concentration $c\sim11.7\pm0.6$.\\
  We further analyse the stellar kinematics data of \citet{sand04} with
  a detailed modelling of the line-of-sight velocity distribution of stars
  in the cD galaxy and quantify the small bias due to non-Gaussianity of the
  LOSVD. After correction, the NFW lens model is unable to properly fit
  kinematical data and is a factor of $\sim2$ more massive than suggested
  by X-rays analysis \citep{allen01}.\\
  The discrepancy between projected (lensing) and tridimensional
  (X-rays,dynamics) mass estimates is studied by assuming prolate (triaxial)
  halos with the major axis oriented toward the line-of-sight. This model
  well explains the high concentration and the misalignement between stellar
  and dark matter components $(\Delta \psi \sim 13^\circ)$.\\
  We then calculate the systematic and statistical uncertainties in the
  relative normalization between the cylindric $M_2(<r)$ and spherical
  $M_3(<r)$ mass estimates for triaxial halos. These uncertainties prevent any
  attempt to couple 2D and 3D constraints without undertaking a complete
  tridimensional analysis. Such asphericity/projection effects should be
  a major concern for comparisons between lensing and X-rays/dynamics
  mass estimates.
}
\maketitle

\section{Introduction}
The issue of the late non-linear evolution of cosmic structure is
essentially addressed via large N-body cosmological simulations.
It is important to test their validity by comparing the
small scale matter distribution to numerical predictions.
Two observations act as key-tests for the CDM paradigm:
the mass distribution of dark matter halos (radial density profile and
triaxiality) and the abundance of sub-halos within main halos. This
work focuses on the former issue.

Most CDM simulations predict a universal profile of the general form:
\begin{equation}\label{eq:univprof}
  \rho(r/r_s) = \rho_s (r/r_s)^{-\alpha} \left(1+r/r_s\right)^{\alpha-3}\,,
\end{equation}
with an inner slope $\alpha$ ranging between $\alpha=1$ and
$\alpha=1.5$ \citep{NFW97,moore98,ghigna00,jing-suto00}.
The parameters $r_s$ and $\rho_s$ can be related to the halo mass
\citep{bullock01,eke01}. Though more recent simulations propose 
a slightly different universal analytical form \citep{stoehr02,navarro04}.

The global agreement between observations and simulations is subject to
controversy. The inner slope of dark matter halos of low surface
brightness (LSB) dwarf galaxies as inferred from rotation curves tends
to favor soft cores with $\alpha\lesssim0.2$
\citep[{\it e.g.}~][]{salucci01,deblok03,gentile04},
leading to the so-called cusp-core
debate. Many observations have focused on LSB galaxies because their
baryonic content can be neglected and the dark matter distribution in
the halo shall match simulations. However, departs from axisymmetry
(triaxial halos) make the interpretation of rotation curves more
complex and could reconcile observations and CDM predictions
\citep{hayashi04}. The question of the very central mass profile on
dwarfs scales is still open... 

Recently a similar discrepancy at clusters of galaxies scales
is claimed by \citet[][hereafter Sa04]{sand02,sand04}. Using HST
images (allowing the modelling of strong gravitational lensing
configurations) together with Keck spectroscopy (providing the radial
velocity dispersion of stars in the central cD galaxy of the cluster,
the BCG) on a sample of six clusters, these authors found that the
inner slope of the dark matter halo must be significantly flatter than
that measured in simulations. Typically, on a subsample of three
clusters with radial arcs, they found an inner slope
$\alpha=0.52\pm0.05$ (68\%CL). This result takes advantage of the
joint constraints provided by lensing and stellar kinematics. However,
the lensing part of the analysis of Sa04 has been independently
discussed by \citet{bartelmann03b} and \citet{dalal03} because they
did not take into account the lens ellipticity when using the
critical lines radii as a constraint on the density profile.
These two latter authors found that the mass profile is consistent
with a NFW model. The analysis of Sa04 couples 2D projected (from
lensing that deal with mass enclosed in the cylinder of radius $R$)
and 3D tridimensional (from stellar dynamics which project an indirect
information on the mass enclosed in the sphere of radius $r$) mass estimates.

Comparing lensing and X-rays cluster mass estimates is another way to
couple 2D and 3D mass constraints. The overall agreement between X mass
and the mass enclosed in the Einstein radius of clusters are been addressed
by various authors \citep{miralda95b,allen98,wu00,arabadjis04,smith05}.
In most cases, slightly depending on the presence of cooling flows or
the degree of relaxation of the cluster, strong lensing mass estimates
are often larger by a factor $\gtrsim1.5$.{\bf With better S/N data, there is
an increasing evidence that the assumption of spherical symmetry starts
being oversimplistic and may play a important role in this systematic trend
\citep{piffaretti03,filippis05,oguri05,hennawi05b}.}

In this paper, we focus on the density profile of the cluster
MS2137-21 which is part of the Sa04 sample and search for further evidence
for triaxiality in this peculiar cluster.
In Sect. \ref{sec:lensmodel} we present the strong and weak lensing
modelling of MS2137 with a NFW model and show that it is consistent
with all the lensing data at hand from 10 kiloparsecs to 1 megaparsec.
In Sect. \ref{sec:velosec} we develop a detailed method for the
analysis of stellar kinematics and apply it to the best fit NFW model
derived in the previous section. We then discuss the overall agreement 
between lensing mass estimates and the constraints from the
stellar kinematics and X-rays observations of \citet{allen01}.
In Sect. \ref{sec:discrep} we investigate the origin on the
systematic overerestimate of lensing mass estimates
as compared to that of 3D analyses, and show that the
tridimensional shape of halos (prolate, triaxial) is likely to explain
such discrepancies. In Sect. \ref{sec:discrepGEN} we calculate the
statistical properties of the relative normalization between 2D and
3D mass estimates of triaxial halos. We discuss our results and conclude in
Sect. \ref{sec:discussion}.

Throughout this paper, we assume a \mbox{$\Omega_m=0.3$},
\mbox{$\Omega_\Lambda=0.7$} and \mbox{$H_0=70 \h\kms\Mpc$},
leading to the scaling \mbox{$1\arcsec=4.59\hmkpc$}.

\section{Lens modelling}\label{sec:lensmodel}
\subsection{Optical data and $\chi^2$ definition}\label{sec:lensobserv}
In this section we focus on the density profile modeling using lensing
constraints only. The lens properties of the cluster of galaxies have been
extensively studied \citep{fort92,mellier93,miralda95,bartelmann96,hammer97,
  gavazzi03,sand02,sand04,dalal03}. The cluster's redshift is
$z_l=0.313$ and both radial and tangential arcs lie at $z_s=1.501$
\citep{sand02}, leading to the critical surface density $\scrit = 2.39
\times 10^{9}\,h_{70}\,{\rm  M}_\odot \kpc^{-2}$.

Our analysis builds on the previous work of \citet{gavazzi03}
(Hereafter G03). More precisely, we use 26 multiple conjugate knots in
the tangential and radial arcs systems. The method and the knots
location are presented in G03.
{\bf Here, we inflate the uncertainties on knot positions in order to
  account for possible bad associations. Basically, the mean positional error
  is raised to the more realistic value $\sigma_x = 0\farcs3$.
  After a more detailed analysis of images, the G03 value $=0.18\arcsec$ turns
  out to be underestimated. Moreover there was a mistake in the calculation
  of error bars for model parameters in this earlier paper.
  The uncertainty on each knot location is just increased by the same amount,
  so we do not expect any change in the best fit model.
  \citet{dalal03} proceeded in the same way by inflating the G03 errors up to
  a value of $1\arcsec$ which is far too much.
  We stress that the error bars of G03 on the best fit parameters are larger
  than the ones we shall present in the following although they assumed
  smaller uncertainties on knots locations. This is a clear evidence for
  an error in the analysis. The present updated results should be considered
  as correct.
}
We also exclude constraints from the fifth
central demagnified image reported in G03 since its detection is
marginal and is not confirmed by Sa04. We use a personal ray-tracing
inversion code which includes many aspects of the {\tt lensmodel} software
\citep{keeton01soft2,keeton01soft1}. In particular, we adopt the same
source plane $\chi_{\rm src}^2$ definition.
 
In addition, we simultaneously include weak lensing constraints also presented in
G03. The catalogue of background ``weakly lensed'' galaxies comes from VLT/FORS
and VLT/ISAAC images for which we were able to derive a good estimate of
photometric redshifts using $UBVRIJK$ bands. We fully compute the likelihood 
as a function of model parameters \citep{schneider00,king01}.
\begin{equation}\label{eq:lik-wklen}
\mathcal{L}_{\rm wl} = \prod_{i=1}^{N_{\rm bg}}
p(e_i)
\end{equation}
where $e_i$ is the observed ellipticity of the background galaxy. We have
\begin{equation}\label{eq:jacob-eies}
  p(e_i)=p_s(e_s(e_i,g_i)) \left\vert \frac{\der e_s}{\der e_i}\right\vert ,
\end{equation}
$e_s$ being the source ellipticity and $g_i=g(\vec{\theta}_i,z_i)$ is
the reduced shear. See \citet{geiger98} for the description of the
relation $e_s(e_i,g)$ and for the corresponding transformation
Jacobian. Ellipticities are measured on the I band image. We
improved the previous analysis of G03 and built a new PSF smearing
correction pipeline based on the KSB method \citep{KSB95} but leading
to a better weighting scheme \citep{gavazzi04}. We fully take into
account redshift information, either photometric\footnote{using {\tt hyperz}
  facilities \citep{Bolzonella00}, see also G03} for weak lensing or
spectroscopic for strongly lensed arcs. The use of photometric redshifts to
select the sample of background galaxies avoids the problem of contamination
by foreground unlensed galaxies \citep{broadhurst05b}. The global $\chi^2$
for lensing is :
\begin{equation}\label{eq:chi2lens}
  \chi^2_\mathrm{lens} = \chi^2_\mathrm{src} - 2 \ln \mathcal{L}_{\rm wl}\;.
\end{equation}
We use the {\tt minuit} library \footnote{\url{http://cernlib.web.cern.ch/cernlib/}}
to minimize this $\chi^2$.
The error analysis is performed using both {\tt minuit} facilities and
Monte-Carlo Markov Chains (MCMC) based on the implementation of
\citet{tereno04}. We chose to use
MCMCs because {\tt minuit} has difficulties to draw $\Delta\chi^2$
contours in the parameter space when there are strong degeneracies. In
order to fasten the convergence of the chains, we previously run many
{\tt minuit} optimizations starting from a broad range of initial
conditions. From the well defined best fit minimum position, we
started up to five chains with each of them ending with $6\times10^4$
relevant iterations. The convergence assessment was done in the same
way as in \citet{tereno04}.

\subsection{Luminosity profile}\label{sec:lumdist}
We fitted the central cD surface brightness on the F702/WFPC2 Hubble Space
Telescope image \citep{hammer97} with a general projected stellar
density profile assuming that all stars have a constant mass-to-light
ratio.
{\bf  We assumed the following analytic
expression for the three-dimensional radial distribution of stars :
\begin{equation}
  \rho_*(r) = \rho_{s*} x^{-\delta} (1+x)^{\delta-4},
\end{equation}
where $x=r/r_{s*}$, $r_{s*}$ is a scale radius. We considered the particular
values $\delta=1$ for the Hernquist profile \citet{hernquist90} and $\delta=2$
for the Jaffe profile \citet{jaffe83} which were added to the 2 dimensional
galaxy model fitting software {\tt galfit} \citep{peng02}.
Before fitting the luminosity profile is convolved by the HST/F702 PSF.}
The Hernquist fit gives an axis ratio $q_*=b/a=0.83\pm0.12$, a scale radius
$r_{s*}=11.1\pm1.9\hmkpc=2\farcs4 \pm 0\farcs1$ and a reduced
$\chi^2/{\rm dof} = 10.2$. This latter value could be noticeably decreased by
taking into account a rotation of the major axis position angle within the inner
3 arcsec (see Fig. 2 of G03). We also tried to fit a Jaffe profile as proposed
by Sa04 but we found a much worse $\chi^2/{\rm dof} = 99.0$.
{\bf However, Sa04 found that the stellar density profile is well fitted by a
  De Vaucouleur model which is bracketed by the Hernquist and Jaffe models.}
Therefore we consider that the stellar component is well modeled by an
Hernquist density profile with $r_{s*}=11.1\hmkpc$ and a total rest frame V
band luminosity \mbox{$L_V = 4.77\pm0.40 \times 10^{11}\,\hmlsun$}
(Sa04). The mass content in stars is 
\mbox{$M_* = 2 \pi \rho_{s*} r_{s*}^3 q_* \equiv \Upsilon_V L_V$} where
$\Upsilon_V$ is the rest-frame V band stellar mass-to-light ratio.

{\bf
  Throughout the paper, we shall discuss the consequences of this particular
  choice. At this level, we expect the effect of this choice to be more
  important for stellar kinematics than for lensing. This can be understood
  because  strong lensing constraints probe the total density profile well
  beyond the stellar scale radius where Hernquist, Jaffe or De Vaucouleur
  profiles are very similar (see Fig. 2 of Sa04).
}

\begin{figure*}[htbp]
  \resizebox{17cm}{!}{\includegraphics[angle=-90]{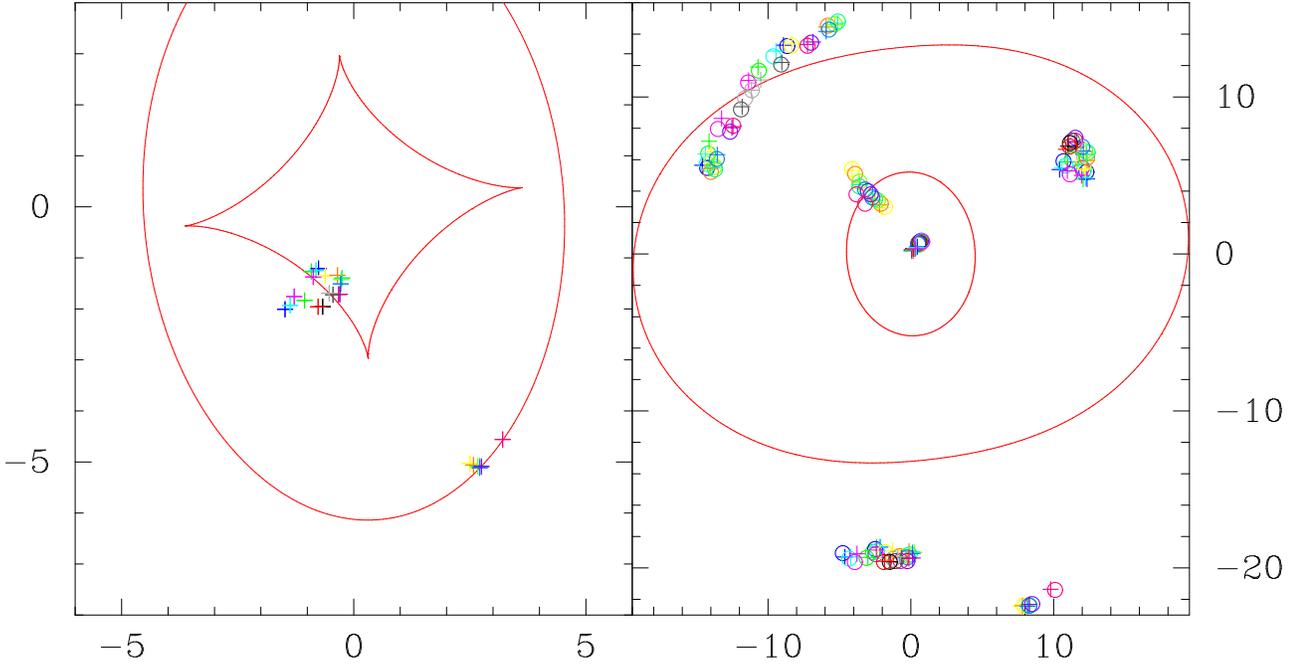}}
  \caption{\scriptsize NFW best fit model for strong lensing constraints.
    {\it \ \ Left:} Caustic lines with
    the position of the 26 knots in the source plane. {\it \ \ Right:}
    Critical lines with the observed (resp. model) position of the 26
    knots represented with circles (resp. + signs). Geometrical units
    are arcsec.
  }
  \label{fig:massprof-nfw}
\end{figure*}

\subsection{NFW dark matter density profile}\label{sec:lensNFW}
The lens is decomposed into two components. The cD stellar component is
modeled by the elliptical Hernquist profile of the previous section.
The stellar mass-to-light ratio $\Upsilon_V$ is treated as a free parameter
with a broad uniform prior $1.5\le\Upsilon_V\le10$ whereas the scale radius,
orientation and ellipticity are fixed by the observed light distribution.
The dark matter halo is modeled with an elliptical NFW density profile: 
\begin{equation}\label{eq:defNFW}
  \rho_{\rm DM}(r) = \rho_s\, (r/r_s)^{-1} \left( 1 + r/r_s\right)^{-2}\,.
\end{equation}
The lens properties of such a density profile are presented in
\citep{bartelmann96}. We used numerical integrations algorithms for
the elliptical\footnote{which are not approximated by elliptical lens
  potentials (numerically faster but leading to unphysical surface
  mass density at large radius)}  Hernquist and NFW density profiles 
\citep{keeton01soft2}. The model has five free parameters :
\begin{itemize}
\item the dark halo scale radius : $r_s$,
\item the characteristic convergence : $\kappa_s \equiv \rho_s r_s/\scrit$,
\item the dark halo axis ratio : $q=b/a$, 
\item the dark halo major axis position angle : $\psi_0$,
\item the stellar mass-to-light ratio : $\Upsilon_V \equiv M_*/L_V $.
\end{itemize}

\reffig{fig:massprof-nfw} shows the best fit NFW strong lensing configuration.
One can see the source and image planes with caustic and critical lines
together with the location of the 26 multiply images knots.
Every observed point (circle) is well reproduced by the model
(+ signs). The source associated to the tangential arc is clearly
crossing the corresponding caustic line (inner astroid) whereas one
can only see the part of the source associated to the radial arc that
is inside the radial caustic (since the part outside the caustic is
not multiply imaged and is useless for modelling). The central image associated
to the tangential system is plotted but is not taken into account in
the modelling. Besides the critical lines location (the only constraint
used by Sa04), our model also remarquably explains the position of
counter-images, the azimuthal configuration, the length and width of arcs.

\begin{table*}[htbp]
  \centering
  \begin{tabular}{ccccc}\\\hline\hline
                & (SL) & (SL+WL) & (1)  &unit\\\hline
    $\kappa_s$   & $0.67\pm0.05$ & $0.66\pm0.03$  & $0.30\pm0.15$  &   \\
    $r_s$    & $158\mypm{15}{13}$ & $162\mypm{11}{9}$ & $160\pm30$ &$\hmkpc$ \\
    $r_{\rm 200}$& $1.88\pm{0.05}$ & $1.89\pm{0.04}$ & $1.39\mypm{0.49}{0.38}$ &$\hmMpc$\\
    $M_{\rm 200}$ & $7.56\mypm{0.63}{0.54}$ & $7.72\mypm{0.47}{0.42}$ & - &$10^{14}\hmmsun$ \\
    $c$          & $11.92\mypm{0.77}{0.74}$ & $11.73\pm0.55$ & $8.7\mypm{1.2}{0.9}$& \\
    $q$          & $0.774\pm0.010$ & $0.777\pm0.007$ & -  &  \\
    $\psi_0$   & $5.86\pm0.14$ & $5.88\pm0.13$ & -  & deg \\
    $\Upsilon_V$ & $2.40\pm0.45$ & $2.48\pm0.39$ & -  &$\mslsun$\\\hline\hline
  \end{tabular}
  \caption{\scriptsize Best fit NFW model parameters and their 68\% CL
    uncertainty (marginalized over all the other parameters).
    (SL) corresponds to a model in which weak lensing constraints
    are ignored whereas (SL+WL) takes both strong and weak lensing
    constraints into account. (1) refers to the CHANDRA
    X-rays values of \citet{allen01}. The apparent disagreement
    between their estimates and ours is discussed in the text.} 
  \label{tab:resNFW}
\end{table*}

The model requires a rest frame V band stellar mass-to-light ratio
$\Upsilon_V=2.5\pm0.4$. This value is in good agreement with
expectations of evolution of $2 \lesssim t\lesssim 4$ Gyr old stellar
populations. {\bf The reason why the stellar mass content is so tightly
constrained is that the stellar and dark matter components are not aligned.
There is a position angle misalignment of $\Delta \psi=13^\circ$.
This was first pointed out by G03. Otherwise, there would be a degeneracy
between the relative contribution of dark matter and stars. Here the
degeneracy is broken though the contribution of stars is subdominant at all
scales (and a factor $\sim 2$ at the centre) as shown in
\reffig{fig:mass-compar}. This situation well explains
the small inaccuracy in the radial arc modelling highlighted
in Sect. 4.2 of G03. By adding a small misaligned contribution under the form
of stars at the center, one is able to twist to isopotentials and precisely
reproduce the radial arc and its counter-image. See \citet{romanowsky98}
and \citet{buote02} for a similar example. We shall turn back to this issue
in Sect. \ref{sec:discrepDISC} and appendix \ref{append:triax}.
It is worth mentioning that changing the stellar mass profile to a Jaffe model
does not make differences. The total (misaligned) stellar mass is well fixed
by lensing.}

\begin{figure}[htbp]
   \resizebox{\hsize}{!}{\includegraphics{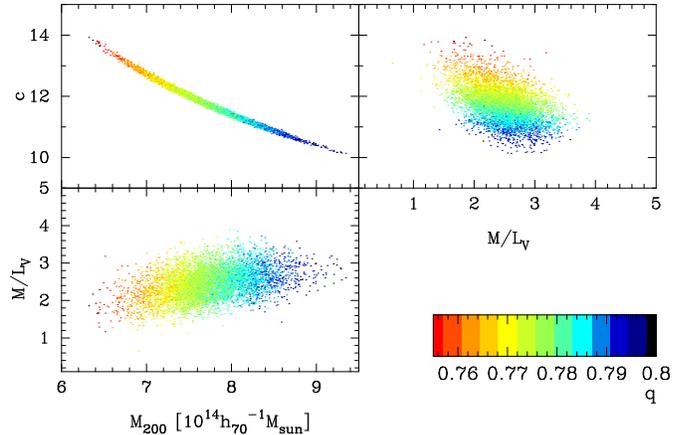}}
  \caption{\scriptsize Scatter plot showing the projection of MCMCs on
    some planes of the parameters space of the NFW lens
    modelling. The color codes for the
    ellipticity parameter according to the scale at the bottom right
    corner. The axes are the virial mass $M_{\rm 200}$, the
    concentration $c$ and the rest-frame V band stellar mass-to-light
    ratio $\Upsilon_V \equiv M_*/L_V $.
  }
  \label{fig:scatterMCMC}
\end{figure}

In Table \ref{tab:resNFW} we present the best fit NFW model parameters
in terms of more physical quantities like the virial radius $r_{\rm 200}$,
the concentration parameter $c=r_{\rm 200}/r_s$ or the virial mass $M_{\rm 200}$
that all derive from $\kappa_s$, $r_s$ and $q$. At the best fit parameter
set, the minimum $\chi^2$ value is $\chi^2/{\rm dof}=
4931.20/4965=0.993$. 
{\bf When considering strong lensing constraints only,
  $\chi^2_{\rm src}/{\rm dof}=76.4/130=0.59$ showing that both
  strong and weak lensing observations are well modeled\footnote{
  If we have used the former positional uncertainties of G03, the best fit
  model would not have been changed but we would have found a minimum
  $\chi^2_{\rm src}/{\rm dof}=1.64$ which is also a acceptable fit.}.}
The (SL) and (SL+WL) columns detail how the best fit model is changed whether
weak lensing constraints are added to the model or not. Basically, errors are
just reduced and no significant change in the best fit parameters value is
observed. \reffig{fig:scatterMCMC} shows the degeneracies between the
concentration parameter, virial mass, stellar mass-to-light ratio and
ellipticity (color-coded).

{\bf \reffig{fig:mass-compar} shows the radial projected mass profile for the best
  fit NFW+stellar components as well as a detail of the stellar component.
  The thickness of curves is representative of the 1$-\sigma$ uncertainties.
  This is done by considering many points of the MCMCs that stand within the 
  1-$\sigma$ contour about the best fit model.
}

\begin{figure*}[htbp]
  \centering
   \resizebox{17cm}{!}{\includegraphics{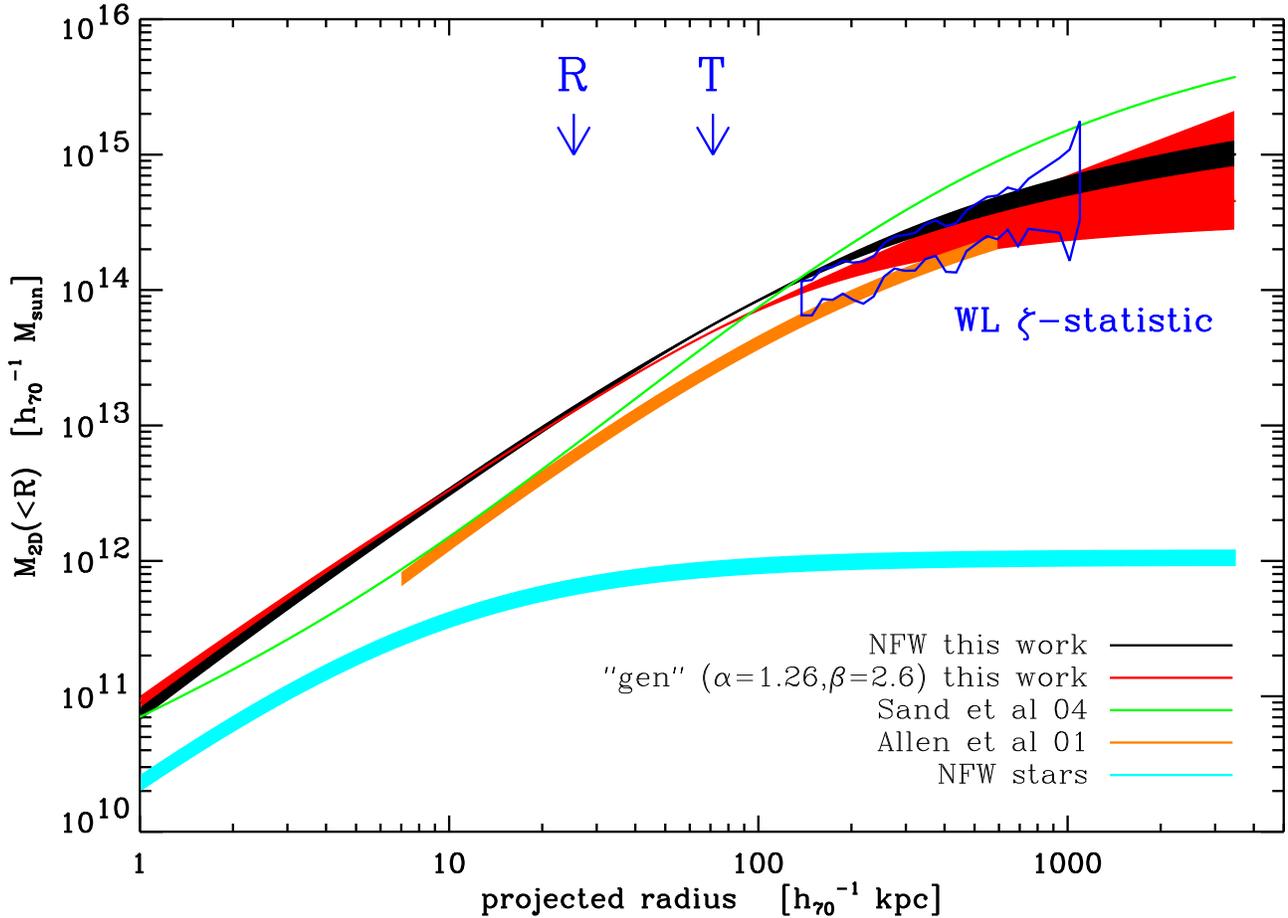}}
  \caption{\scriptsize Projected total mass (stars+DM) profile within
    cylinder of radius $R$ for the best fit NFW model (black), the
    best fit general {\it ``gen''} model with $\alpha=1.26$ and $\beta=2.6$
    (red), the best fit model of \citet{sand02,sand04} (green) and the
    best fit NFW model from X-rays constraints \citep{allen01} (orange).
    The contribution of stars for the NFW model is the cyan dashed curve.
    The width on the curves is representative of the 1$-\sigma$ uncertainties
    (except for the Sa04 profile).
    The blue region is the domain constrained by the weak-lensing
    $\zeta$-statistic. The location of the tangential (resp. radial) critical
    lines is reported by a blue ``T'' (resp. ``R''). The {\it ``gen''} and NFW
    models match well from the center to the inner radius probed by
    weak-lensing. Beyond, their are consistent within the weak-lensing
    uncertainties.
    For $R\lesssim40\kpc$ both profiles present large (a factor $\sim2$)
    discrepancies with the curves inferred by Sa04 and Al01. {\it See text}. 
  }
  \label{fig:mass-compar}
\end{figure*}

\subsection{A more general profile for dark matter}\label{sec:lensGEN}
The quality of the fit is such that very few departs from the NFW model 
we found are allowed. In order to check how restrictive the analytical
form \eqref{eq:defNFW} is, we also assumed the following profile for the
dark matter component \citep{wyithe01}
\begin{equation}\label{eq:rhocusp}
  \rho_{\rm DM}(r) = \rho_s\, x^{-\alpha} \left(1+x^2\right)^{(\alpha-\beta)/2}\:,
\end{equation}
with $x=r/r_s$ the radius in units of a scale radius $r_s$ and two more free
parameters: the asymptotic inner and outer slopes $\alpha$ and $\beta$
respectively. We will refer to this model as the \gen profile.
{\bf This model slightly differs from the generally assumed generalised
  gNFW model $\rho \propto x^{-\alpha} \left(1+x\right)^{\alpha-3}$ and a 
  comparison to previous studies is not straightforward. Nevertheless
  we chose this model because it is computationally tractable even 
  with elliptical symmetry \citep{chae98,chae02} and allows another
  degree of freedom since the outer slope is not fixed.}

{\bf For the best fit model, we have $\chired=4915.90/4963=0.990$, and
  $\chi^2_{\rm src}/{\rm dof}=64.2/128=0.50$. Here again, the $\chi^2$ value
  is satisfying\footnote{It would have increased to $\chi^2_{\rm src}/{\rm dof}=1.39$
  using the G03 uncertainties.}.}
We found $\Upsilon_V=2.09\pm0.16$ also consistent with
stellar evolution models, $\beta=2.69\mypm{0.32}{0.22}$ and
$\alpha =1.262\mypm{0.013}{0.017}$.The constraints on $\alpha$ are very tight and
show that lensing is inconsistent with any soft core $\alpha \ll 1$.
However, it does not contradict the NFW behavior $\rho\propto r^{-1}$ at
small scales because the fast transition $1+x^2$ in the \gen profile
differs from the NFW case $(1+x)$. This can clearly be seen in
\reffig{fig:mass-compar} where the projected NFW and \gen mass profiles
match over a broad radius range ($r\lesssim100\kpc$). The differences at
larger scales are still within the weak lensing uncertainties.
Consequently, we can faithfully trust the radial behavior of the lensing
deduced mass profile of the NFW model between $10<R<1000\kpc$.

{\bf Here again, changing the Hernquist stellar profile to a Jaffe model does
  not change our results.}

\subsection{Comments}\label{sec:lensCOM}
\begin{figure}[htbp]
  \resizebox{\hsize}{5cm}{\includegraphics{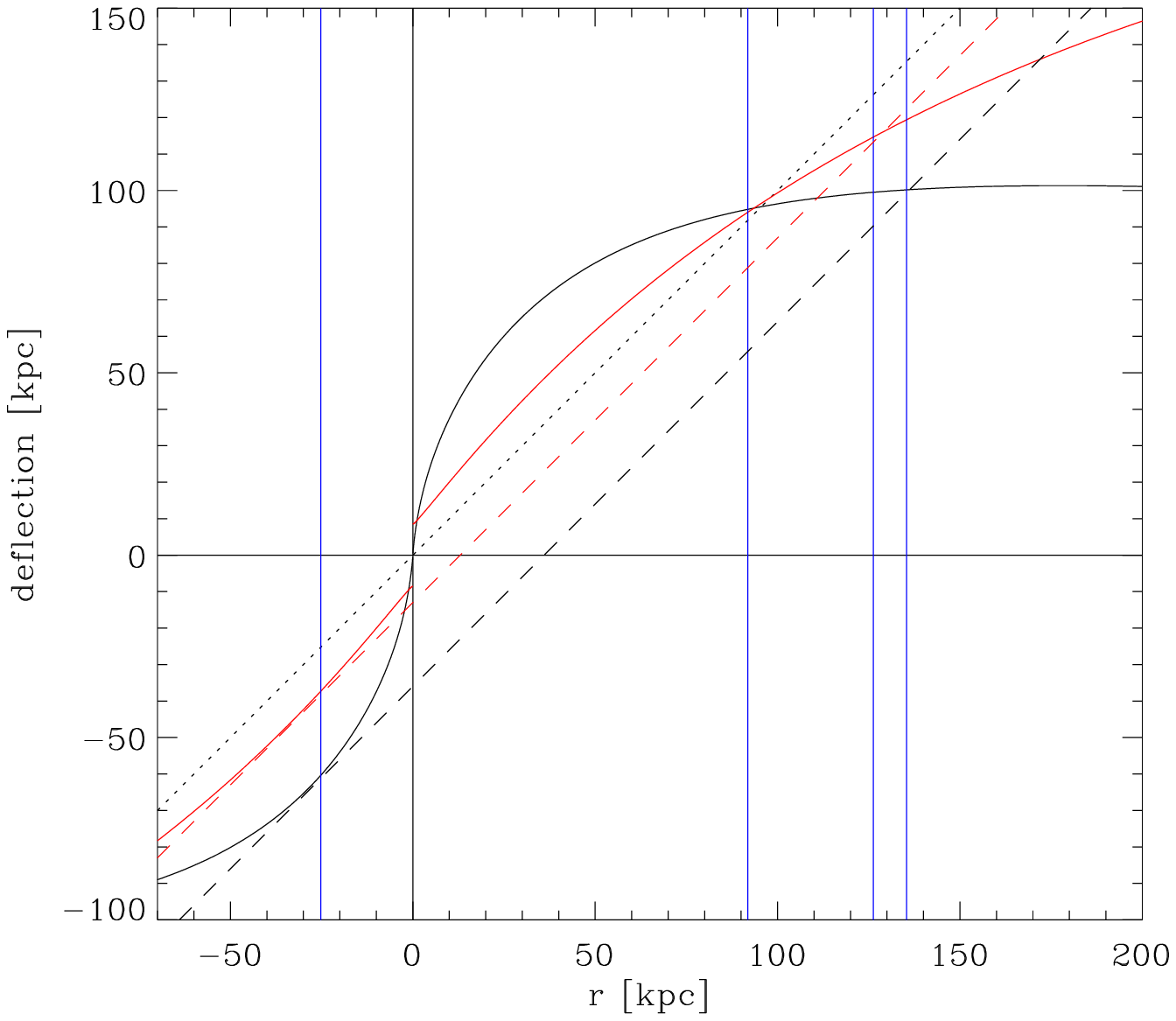}}
  \resizebox{8.5cm}{5cm}{\includegraphics{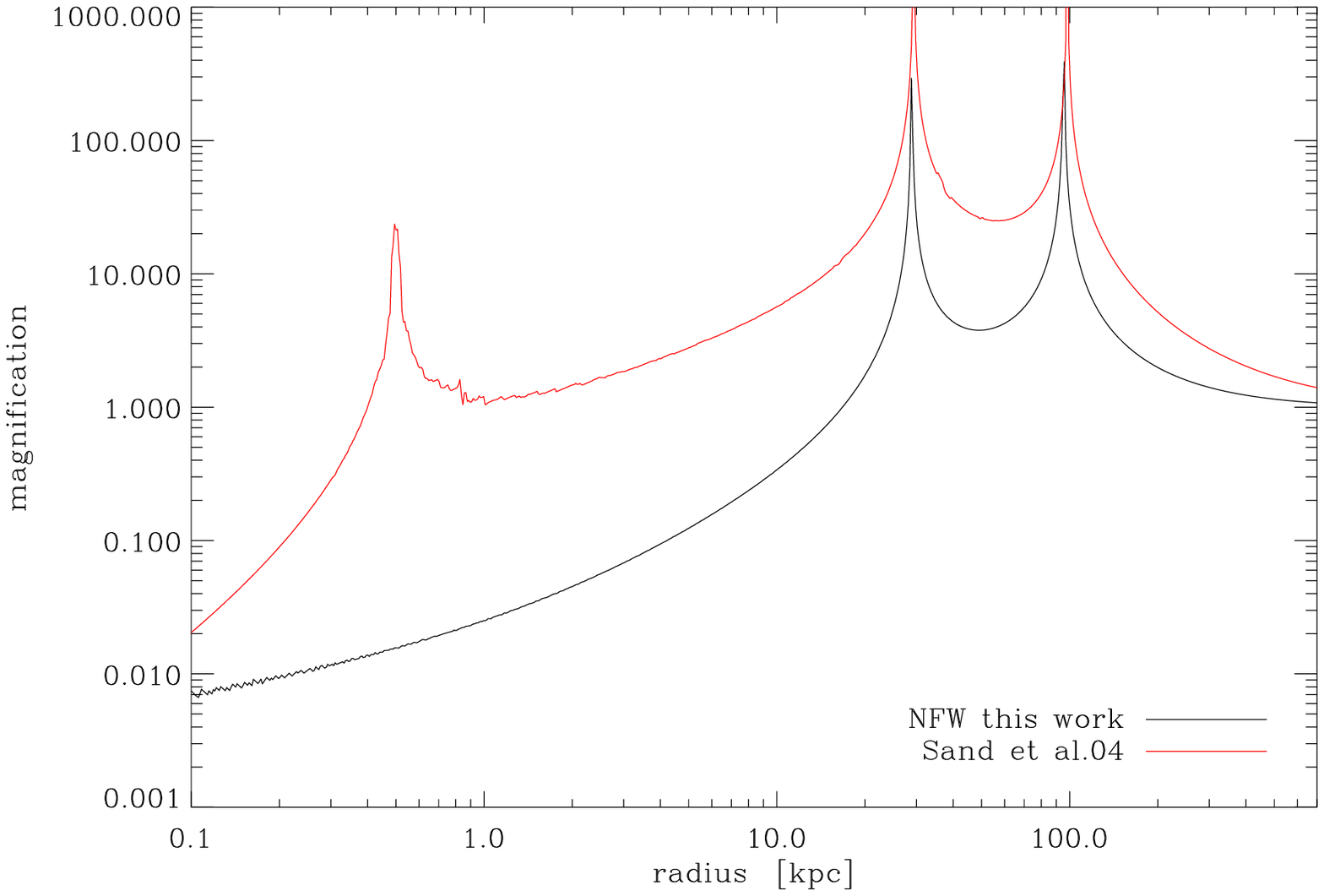}}
  \caption{\scriptsize {\it Upper panel:} Circularly averaged deflection
    angle for our best NFW fit (solid black curve) compared to that of
    Sa04 (solid red curve). From left to right, the blue vertical lines
    represent the radial and tangential critical radii, the Sa04 and our
    prediction of the radial arc counter-image. The critical radii match
    from one model to another but the radial arc lenght significantly differs.
    The dotted $y=r$ line gives the solution of the
    tangentiel critical radius. The dashed lines $y=r+u$ (with $u$ the source
    location) give the solution of the radial critical radius where it is
    tangent to the curve $y=r$. {\it Lower panel:} Magnification profile. 
    The Sa04 model predicts high magnifications close to the center with
    another radial critical line at the very center ($r\sim0.1\arcsec$).
  }
  \label{fig:lens-compar}
\end{figure}

\reffig{fig:mass-compar} also shows the best fit model of Sa04 which presents
strong discrepancies with both our NFW and \gen models. Though the projected mass
at the tangential arc radius ($\sim100\kpc$) matches our estimates, the model
of Sa04 is inconsistent with most lensing constraints. They imposed the radial
critical line to fit the observations but their model cannot reproduce the radial
arc length and its counter-images well, nor the tangential arc width and weak
lensing at $R\gtrsim 200\kpc$. This can be understood by comparing the circularly
averaged deflection $\vec{\alpha}$ profile of these models in the upper
panel of \reffig{fig:lens-compar}. This plot is used to solve the lens equation
graphically. The tangential critical radii (intersection of curves $\alpha(r)$
and $y=r$) are consistent from one model to another. As well the curves
$\alpha(r)$ are tangent to the line $y=r+u$ at the same radial critical radius.
But the intersections of $\alpha(r)$ and this line at the opposite side,
which give the location of the counter-image of the radial arc, significantly
differ ($\sim$ 2 arcsec). Moreover, we can see that the Sa04 model predicts
a much more elongated radial arc that could extend very close to the lens center.
This is clearly excluded by the data. It is worth noticing that the Sa04 model
predicts another radial critical line at the very center ($r\sim0.1\arcsec$)
\footnote{which makes the issue of the fifth image worth observationaly
  addressing.} and globally higher magnifications since $\alpha(r)$
is close to the bisectrix $y=r$. The only models consistent with all
lensing constraints are the ones similar to the NFW and the \gen models.

The column (1) of Table \ref{tab:resNFW} also resumes the NFW model parameters
deduced from Chandra X-rays observations of \citet[][hereafter Al01]{allen01}.
The projected mass profile of their model is the orange thick curve on
\reffig{fig:mass-compar}. This NFW profile is twice as low as our NFW and
\gen models over a range $10<R\lesssim200\kpc$ ({\it i.e.} the factor $1/2$ in
the value of $\kappa_s$ in table \ref{tab:resNFW}).
At larger scales $R\gtrsim300\kpc$, the Al01 mass profile becomes consistent
with weak lensing and our models. X-rays (Al01) and stellar dynamics (Sa04) mass
estimates agree at small scales $R\lesssim50\kpc$.

\section{Dynamics of stars in the BCG}\label{sec:velosec}
The kinematical properties of stars in the central cD galaxy are
studied in this section. Instead of using the standard Jeans equation to
relate the gravitational potential and the velocity dispersion of tracers,
we fully calculate the line-of-sight velocity distribution LOSVD via a
thorough dynamical analysis which is detailed in appendix \ref{append:dynam}.
By doing so, we can estimate the biased velocity dispersion profile
$\vhalf(R)$ in place of the true velocity dispersion of stars $\slos(R)$
due to the assumed gaussianity of absorptions lines.

\begin{figure}[htbp]
  \resizebox{\hsize}{!}{\includegraphics{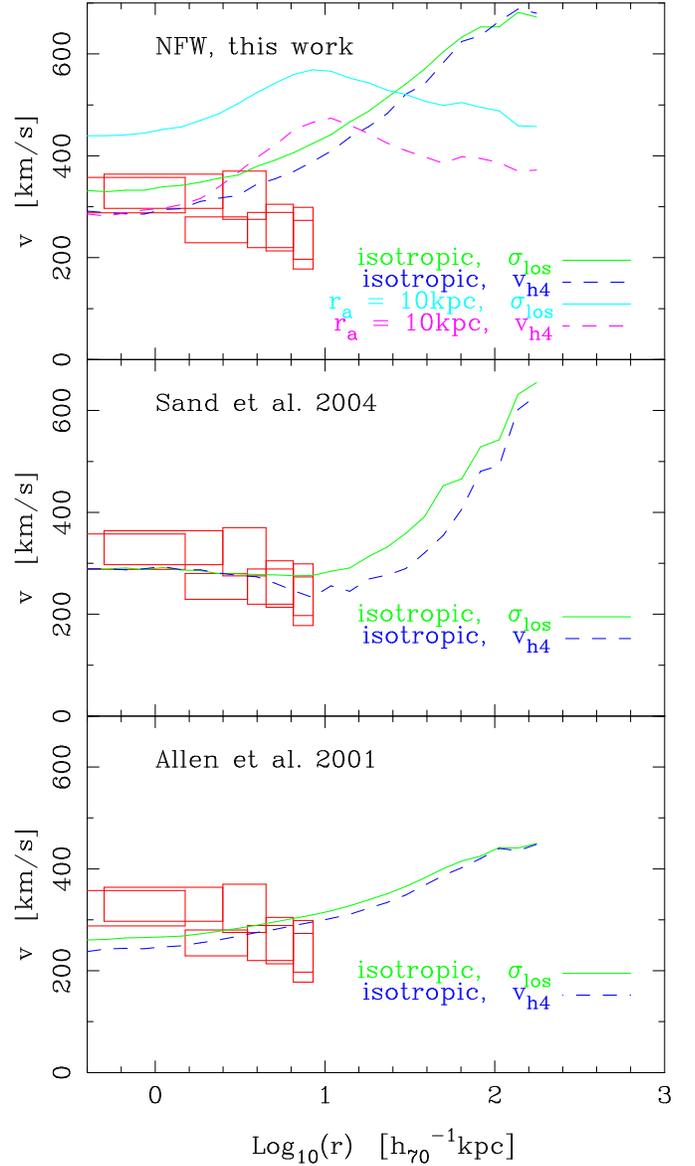}}
  \caption{\scriptsize
    Velocity dispersion $\slos(R)$ (solid green line) and biased velocity
    $\vhalf(R)$ (dashed blue line) profiles for isotropic orbits.
    The measurements of Sa04 are represented by red boxes.
    \ \ {\it Top:} Our best fit NFW model. In this panel we have also 
    represented another couple of ($\slos$,$\vhalf$) curves (cyan and
    magenta respectively) which correspond to an anisotropic Osipkov-Merritt
    case with $r_a=10\kpc$. The introduction of anisotropy does not improve
    the fit quality but leads to huge departs between $\slos$ and $\vhalf$.
    \ \ {\it Middle:} Best fit model of Sa04 which matches the data well.
    \ \ {\it Bottom:} Best fit model of Al01, inferred from X-rays analysis.
    This model also matches the measurements of Sa01 whereas none of these
    latter profiles are consistent with lensing observations.
  }
  \label{fig:velopdf-compar}
\end{figure}

The analysis presented in appendix \ref{append:dynam} shows that departs from
gaussianity are kept at a low level for the lensing-deduced NFW mass model.
For isotropic orbits Gaussianity is a fair assumption:
$(\vhalf-\slos)/\slos\sim -13\%$ at $R\sim1\kpc$ and then decreases whereas
departs can reach $\sim30\%$ for anisotropic orbits. With this mass model
we plot $\slos(R)$ and $\vhalf(R)$ on the top panel of 
\reffig{fig:velopdf-compar} for different values of the anisotropy radius
$r_a=\infty$ and $r_a=10\hmkpc$. The agreement between the measurements
of Sa04 and $\vhalf(R)$ is better than with $\slos(R)$ but introducing
anisotropy cannot improve the fit quality for $R\simeq 10 \kpc$:
the $\vhalf$ curve of the NFW model raises too fast whereas data indicate
a declining tendency. However, if kinematical data would extend to slighly
larger scales, we expect the profile to start raising and get closer to the
model beyond a few tens of kpc as observed in others cD galaxies
\citep{dressler79,kelson02}.

We attempted to couple lensing and kinematical constraints by minimizing
the merit function 
$\chi^2_\mathrm{tot}=\chi^2_\mathrm{lens}+\chi^2_\mathrm{kin}$, with
\begin{equation}\label{eq:chi2kin}
\chi^2_{\rm kin} = \sum_i^{N_{\rm bins}} \frac{ ({\vhalf}_{,i}-v_{\rm mes,i})^2}{\sigma_{\rm mes,i}^2}\:,
\end{equation}
that accounts for kinematical constraints and $\chi^2_\mathrm{lens}$ 
defined in Eq. \eqref{eq:chi2lens}. $v_{\rm mes,i}$ and $\sigma_{\rm mes,i}$
are the measurements and errors of Sa04, and $N_{\rm bins}=8$. The inferred
NFW model is marginaly changed as compared to the one found using lensing only.
The NFW model is overconstrained by lensing and cannot fit the Sa04 kinematical
data better. We find $\chi^2_{\rm kin}/N_{\rm bins}\sim7$ for the best
model\footnote{We find $\chi^2_{\rm kin}/N_{\rm bins}\sim8$ if we change
  $\vhalf$ by $\slos$ in Eq. \ref{eq:chi2kin}, {\it ie} if we neglect the
  velocity bias due to non-gaussian LOSVD. Consequently, the correction has
  a weak effect on the modelling.}.
After $\chi^2$ minimization, the NFW model is still unable to reproduce
the velocity decline at $R\gtrsim4\kpc$.

{\bf 
Changing the Hernquist stellar light profile by a Jaffe model as proposed
by Sa04 slightly improves the fit of kinematical data without altering the
lens modelling (see Sect. \ref{sec:lensNFW}). In this case, we have
$\chi^2_{\rm kin}/N_{\rm bins}\sim6.5$. The velocity dispersion curve raises a bit
slower as compared to the Hernquist case. However, since the mass budget is
dominated by dark matter, there is not much improvement. Lensing constraints
are so tight that the allow region in the parameter space is completely fixed.

Likewise the \gen model also fails in reproducing kinematical data although
it has more free parameters. In this case, the inner slope $\alpha=1.250\pm0.011$
is still fixed by lensing. The inferred stellar mass-to-light ratio is
$\Upsilon_V=1.83\pm0.14$ which is a rather low value. For the \gen profile
too, switching the stellar mass profile to a Jaffe model does not
significantly improve the fit on kinematical data.
}

We have shown that departs from gaussian absorption lines induce a small
bias which starts being important for dynamical systems with radial orbits.
However this bias is unable to explain the discrepancy between lensing and
kinematical mass estimates. Furthermore, such a bias cannot be advocated to
explain the discrepancy between lensing and X-rays mass estimates.
We can see on the central and bottom panels of \reffig{fig:velopdf-compar}
that the mass model of Sa04 fairly reproduces kinematical data, as well as
the Al01 model (provided one addes the contribution of a central cD galaxy
with $\Upsilon_V\sim2.5$).

\section{Discrepancies between mass estimates}\label{sec:discrep}
At this level, let us resume the main problems that arise from the previous
sections. A detailed lens modelling predicts a robust projected mass
distribution that is consistent with NFW universal profiles. We have used
a more general density profile for the dark matter halo in order to check
that any other realistic mass profile should match our best fit NFW model
over a broad range $20\lesssim R\lesssim1000\kpc$. This family of models
turns out to be inconsistent with the X-rays and kinematical mass estimates
that are basically indirect measurements of the 3D mass within radius $r$.
These two latter estimates are mutually consistent for $R\lesssim50\kpc$.

{\bf
Since lensing is sensitive to the integrated mass contrast along the line
of sight, it is natural to expect overestimates due to fortuitous alignments
with mass concentrations which are not physically related to the main halo
of interest. Likewise, departs from spherical symmetry are observed
in N-body simulations \citep[{\it e.g.}~][]{jing02tri} and may bias
lensing estimates. This question has been addressed by various authors
\citep{bartelmann95b,cen97,reblinsky99,clowe04,wambsganss04b,hennawi05b}.
Conclusions about the importance of unrelated structures
(large scale structure LSS) slightly differ from one author to another.
\citet{hoekstra03} found LSS to add noise to mass estimates on large scales
but do not lead to biased estimates since on very large scales the skewness
of the density field is negligible and light rays cross overdense regions
as well as underdense ones. At smaller scales, this becomes obviously wrong
and one expects fortuitous alignments of halos to modify the properties of
halos. \citet{wambsganss04b} claim that such effects can increase
the lensing mass of $\sim 30-40\%$ of halos by a factor of
$\sim 15-20\%$ whereas \citet{hennawi05b} found this effect to change the
lensing cross-sections of clusters by a smaller amount ($\lesssim 7\%$).
See also \citep{hamana04} and \citep{hennawi05} for a discussion of
projection effects on weak lensing cluster surveys.

On smaller scales, \citet{metzler01} found the mass of surrounding
(sub)structures like filaments to add a significant contribution to the
total convergence of a cluster-size lens whereas \citet{clowe04} showed
that triaxiality is an important issue for lensing mass estimates.
In the following, we shall focus on this paticular aspects which has been
found to be important for lensing \citet{oguri03,oguri04} and/or
X-rays observations \citep{piffaretti03,filippis05,hennawi05b}.
}

For a triaxial or oblate/prolate halo, the ratio of the mass enclosed
in the cylinder of radius $R$ to the mass enclosed in the sphere of same
radius will differ from that of a spherically symmetric situation.
In order to illustrate projection effects, we consider an axisymmetric
(either oblate or prolate) NFW density profile of the form :
\begin{equation}\label{eq:nfw-prol}
    \rho(m)= \frac{1}{m (1+m)^2} \quad\mathrm{, with}\;\; m^2 = R^2 + \frac{z^2}{q^2}\;.
\end{equation}
The line of sight is along the $z$-axis and matches the major axis 
of a prolate halo when $q>1$ or the minor axis of an oblate halo when $q<1$.
We can express $m\equiv r \lambda(q,\mu)$ with $\mu=\cos\theta$ and
$\lambda^2=1+\mu^2(1/q^2 -1)$.
{\bf Numerical simulations predict triaxial halos with minor axis
  and intermediate axes $c$ and $b$ with a distribution given by
  \eqref{eq:AXpdfJing}. With these relations, we can numerically
  calculate the distribution of $q$ which is close to gaussian by
  approximating $q=c/\sqrt{b}=0.62\pm0.12$
  (resp. $q=1/\sqrt{bc}=1/0.64\pm0.25$) for an oblate (resp. prolate) halo.
  This is a rough approximation since realistic triaxial halos are not
  systematically aligned with the line of sight but this gives an idea
  of acceptable values of the axis ratio $q$.
}.

Since we are interested in ratios between
mass estimates we pay no attention to normalization constants and write
the exact mass $M_{\rm true}(r;q)$ enclosed by the sphere of radius $r$ as
 \begin{equation}
     M_{\rm true}(r;q) =  \int_0^1 \frac{\der\mu }{\lambda^3} M_{\rm
       true}(\lambda r;1)\;,\label{eq:nfw-3dmass} 
\end{equation}
where $M_{\rm true}(r;1)=\ln(1+r)-\frac{r}{1+r}$ for a NFW profile.
We now calculate the mass $M_{\rm lens}(r;q)$ (resp. $M_{\rm kin}$, $M_{\rm X}$)
as it would be found from lensing (resp. stellar kinematics, X-rays)
measurements performed assuming spherical symmetry.

\subsection{Projection effect on lensing}\label{sec:discrepLENS}
Since lensing measures the projected mass along the line of sight and
owing to the fact that the major/minor axis is aligned, the net effect
of asphericity is to multiply the surface mass density by $q$.
So, we can write
\begin{equation}\label{eq:lensMASS}
  M_{\rm lens}(r;q) = q M_{\rm true}(r;1)\;.
\end{equation}
We plot the ratio $M_{\rm lens}(r;q)/M_{\rm true}(r;q)$ as a function
of radius for various values of the axis ratio $q$ on the top panel
of \reffig{fig:mass-ratio}. We can observe strong discrepancies for extreme
values of the axis ratio. Departs between lensing and true masses tend to
vanish at large scale. Thus, one expects a lower effect of asphericity on
weak lensing based mass estimates.
 
\subsection{Projection effect on kinematics}\label{sec:discrepDYN}
Regarding stellar kinematics, projection effects are much more complex 
because stars are expected to move faster along the major axis and boost
the mass estimate. Assuming a distribution function of stars of the
form $f(E,L_z)$ and a reduced gravitational potential $\Psi(R,z)$
the Euler--Jeans equations read :
\begin{subequations}
  \begin{align}
    \frac{1}{\nu} \partial_R (\nu \overline{v_R^2}) + \frac{\overline{v_R^2}-\overline{v_\phi^2}}{R} = \partial_R \Psi\\
    \frac{1}{\nu} \partial_z (\nu \overline{v_z^2}) = \partial_z \Psi\;.
  \end{align}
\end{subequations}
\citep{chandrasekhar60,hunter77,binney87}. Thus, we can express the
components of the velocity ellipsoid:
\begin{subequations}
  \begin{align}
    \nu \overline{v_R^2} = \nu \overline{v_z^2} = & -\int_z^\infty \der z'\, \nu \partial_z \Psi\\
    \nu R \Omega^2 =  & -\int_z^\infty \der z'\,\left[ \partial_R\nu\partial_z\Psi-\partial_z\nu\partial_R\Psi\right] \;.
  \end{align}
\end{subequations}
with $R^2\Omega^2=\overline{v_\phi^2}-\overline{v_R^2}$. The main difficulty
is to compute the potential and its derivatives generated by an ellipsoidal
distribution of mass $\rho(m)$. To do so we use the formalism of
\citet{chandrasekhar69} \citep[see also][]{qian95} :
\begin{subequations}
  \begin{align}
    \Psi(R,z)-\Psi_0 = & - 2 \pi G q \int_0^\infty \frac{\der u}{\Delta(u)} \int_U^\infty \der m\,m \rho(m)\label{eq:psiRZ} \\
    \partial_R \Psi= & - 2 \pi G q R \int_0^\infty \frac{\der u}{\Delta(u) (1+u)} \rho(U)\\
    \partial_z \Psi = & - 2 \pi G q z \int_0^\infty \frac{\der u}{\Delta(u) (q^2+u)} \rho(U)\;,
  \end{align}
\end{subequations}
with $\Psi_0$ the central potential (which is not relevant for our purpose),
$\Delta(u)=(1+u)\sqrt{q^2+u}$ and $U^2=\frac{R^2}{1+u}+\frac{z^2}{q^2+u}$.
For simplicity we assume that the density of tracers $\nu(R,z)$ does not
contribute to the potential (massless). We also assume that the density of 
tracers is ellipsoidal $\nu(R,z)=\nu(m)$ with the same axis ratio as the dark
halo.

We now calculate the observable luminosity-weighted line-of-light
velocity dispersion
\begin{equation}
  I\slos(R) = \int_{-\infty}^\infty \der z\, \nu(R,z)  \overline{v_z^2}\,.
\end{equation}
If this quantity is assumed to be due to a spherically symmetric system and
is deprojected according to 
\begin{subequations}
  \begin{align}
    \nu(r) =& -\frac{1}{\pi} \int_r^\infty \frac{\der I(R)}{\der R}\frac{\der R}{\sqrt{R^2-r^2}}\\
    \nu\overline{v_r^2}(r) =& -\frac{1}{\pi} \int_r^\infty \frac{\der I\slos(R)}{\der R}\frac{\der R}{\sqrt{R^2-r^2}}\;,
  \end{align}
\end{subequations}
one will calculate a biased radial velocity dispersion $\overline{v_r^2}$.
The corresponding biased mass profile $M_{\rm kin}(r)$ is given by the standard
Jeans equation (with isotropic velocity tensor):
\begin{equation}
  M_{\rm kin}(r) = -\frac{\overline{v_r^2} r}{G} \frac{\der \ln \nu \overline{v_r^2}}{\der \ln r}\;.
\end{equation}
We plot the ratio $M_{\rm kin}(r;q)/M_{\rm true}(r;q)$ as a function
of radius for various values of the axis ratio $q$ on the middle panel
of \reffig{fig:mass-ratio}. Here again projection effects can be huge for
extreme values of $q$. Unfortunately, $M_{\rm kin}(r;q)$ depends on the
profile of tracers $\nu(r)$ and the details of this figure cannot be
representative of a general oblate/prolate NFW halo. For instance the bump
at $r\lesssim1$ is due to our assumed density of tracers which here
corresponds to our model of MS2137. However, departs between $M_{\rm kin}(r;q)$
and $M_{\rm true}$ are always important for high $q$ or $1/q$.

\subsection{Projection effect on X-rays}\label{sec:discrepXRAYS}
Similarly, we calculate the perturbation of asphericity on X-rays mass
estimates. As compared to lensing or dynamics the effect is expected to
be weaker since the gravitational potential is systematically rounder than
the mass. For simplicity we assume that the gas (with density $\rho_g$)
is isothermal and in hydrostatic equilibrium.

\begin{figure}[htbp]
  \resizebox{\hsize}{!}{\includegraphics{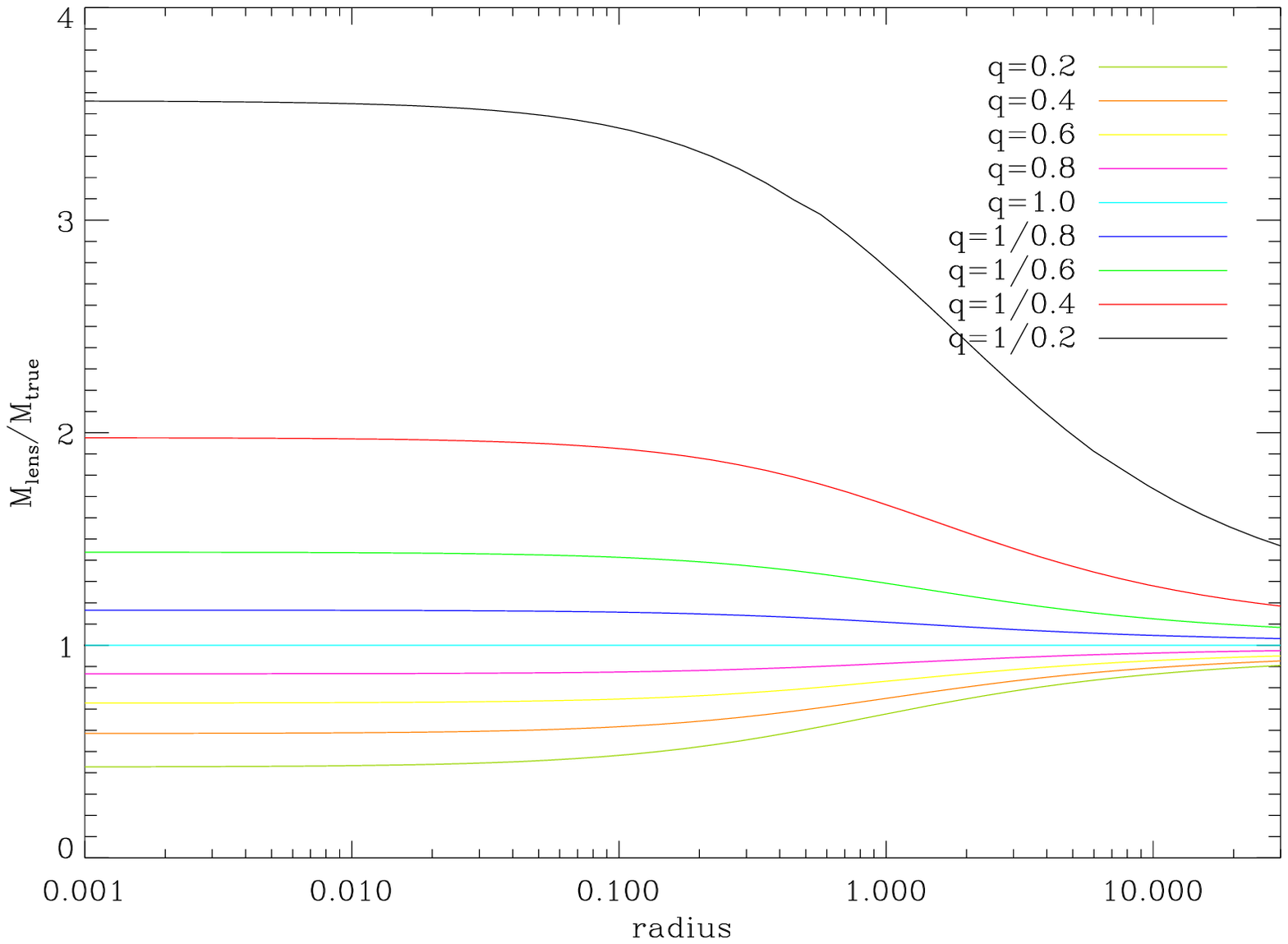}}
  \resizebox{\hsize}{!}{\includegraphics{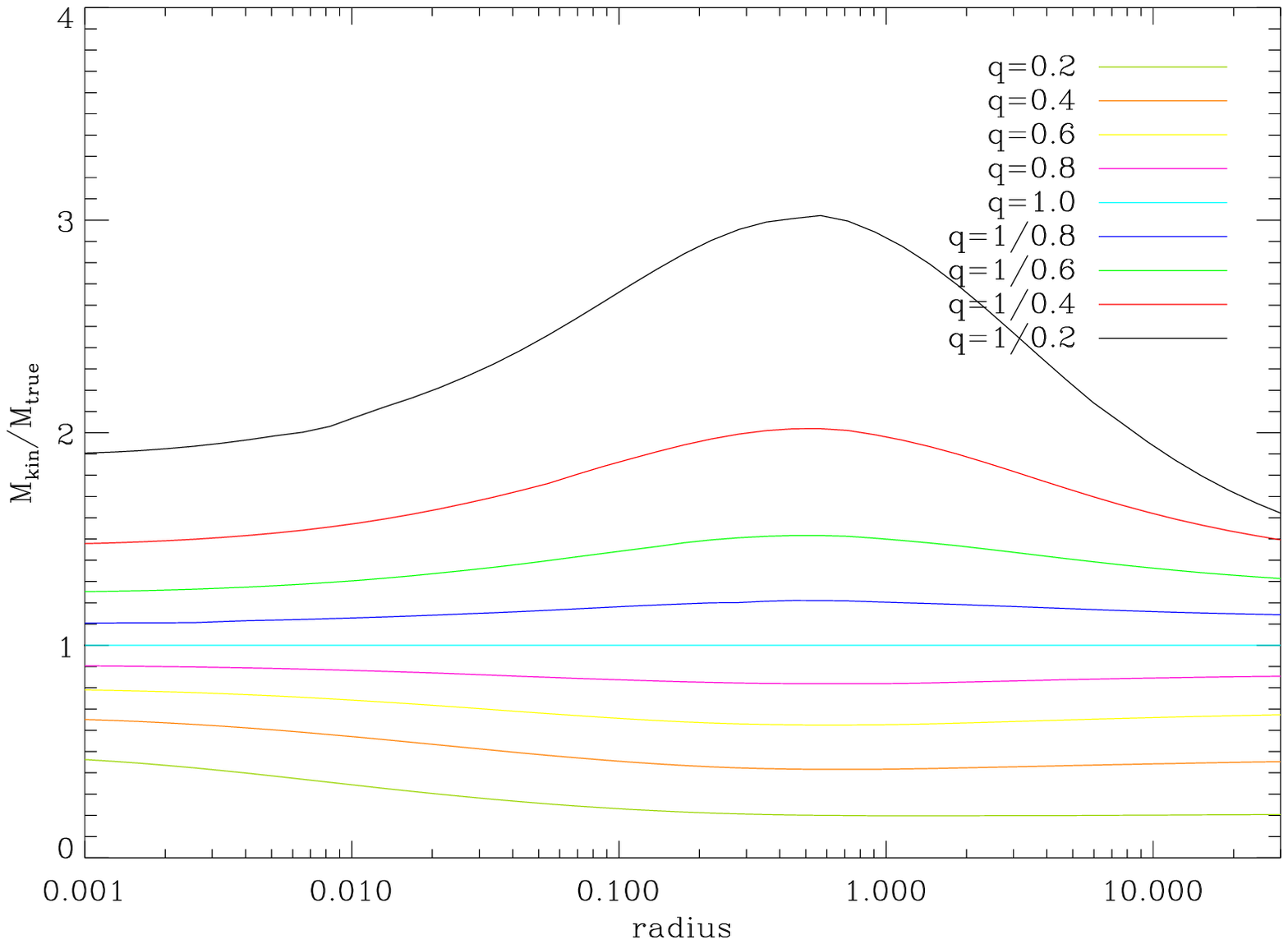}}
  \resizebox{\hsize}{!}{\includegraphics{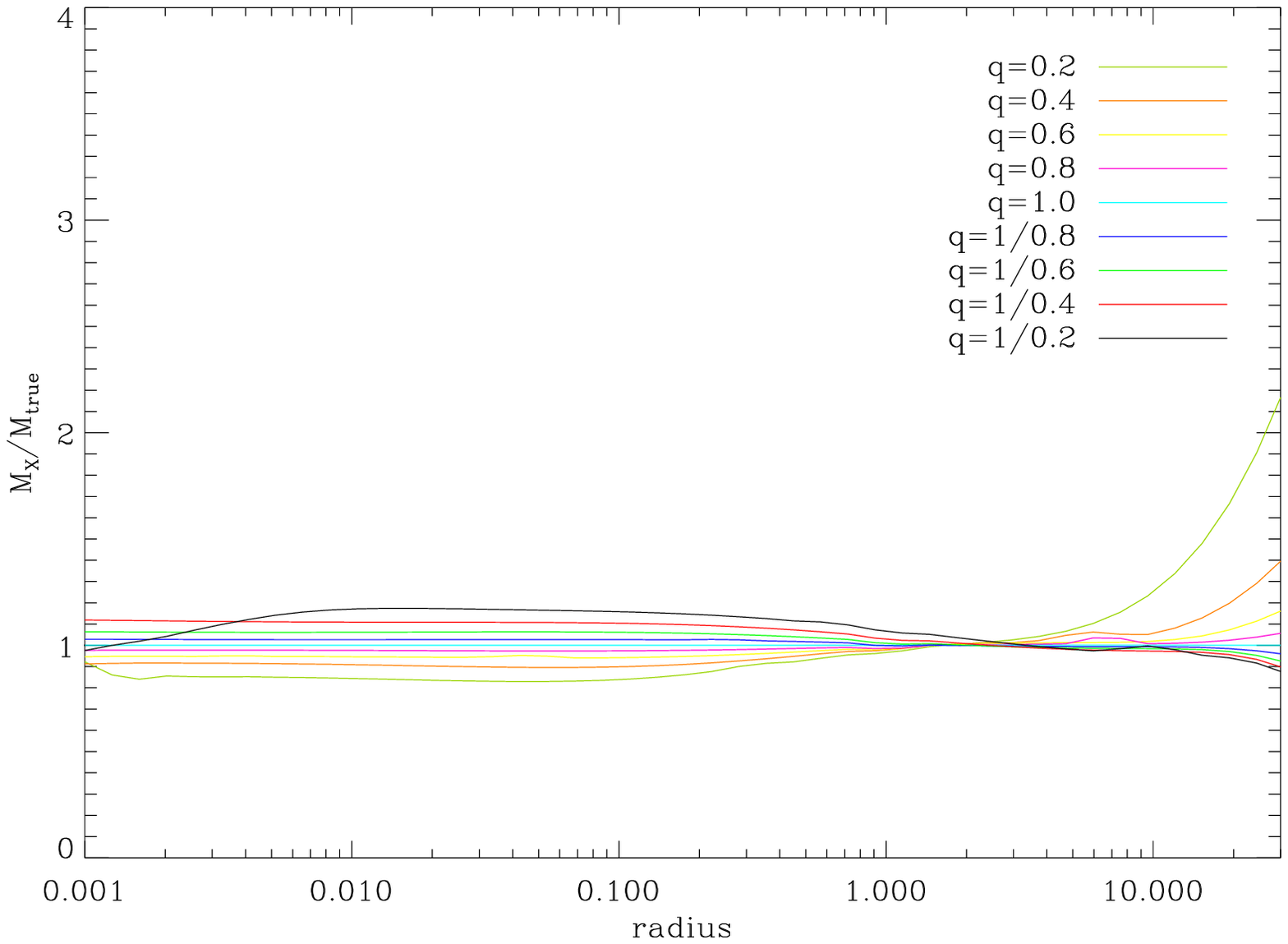}}
  \caption{\scriptsize Radial behavior of ratio between mass estimates for
    various values of the axis ratio
    $q=0.2,0.4,0.6,0.8,1.,1/0.8,1/0.6,1/0.4,1/0.2$.\ \ \ {\it Top Panel:} 
    Ratio $M_{\rm lens}(r;q)/M_{\rm true}(r;q)$. \ \ \ {\it Middle Panel:}
    Ratio $M_{\rm kin}(r;q)/M_{\rm true}(r;q)$. \ \ \ {\it Bottom Panel:}
    Ratio $M_{\rm X}(r;q)/M_{\rm true}(r;q)$. In this latter case, one can
    see much lower values whatever the axis ratio. The asymptotic behavior
    is slightly affected by numerical instability for $r\gtrsim 10$ and
    extreme values of $q$.
  }
  \label{fig:mass-ratio}
\end{figure}

We have $\rho_g \propto \exp[-\Psi/V_0^2]$ with $V_0^2=kT/\mu m_p$.
The X-rays surface brightness of the optically thin gas distribution is 
\begin{equation}
S_X(R) \propto \int \exp[-2\Psi(R,z)/V_0^2] \der z\,,
\end{equation}
with $\Psi(R,z)$ given by \eqref{eq:psiRZ}. Here again, when interpreting
this surface brightness distribution as arising from a spherically symmetric
system, one will deproject $S_X(R)$, obtain a biased gas density
$\rho_g$ and use it in the following equation to obtain the biased
mass profile $M_{\rm X}(r)$.
\begin{equation}
  M_{\rm X}(r)= - \frac{V_0^2 r}{G} \frac{\der \ln \rho_g}{\der \ln r}\,.
\end{equation}
We plot the ratio $M_{\rm X}(r;q)/M_{\rm true}(r;q)$ as a function
of radius for various values of the axis ratio $q$ on the bottom panel
of \reffig{fig:mass-ratio}. As expected $M_{\rm X}(r;q)/M_{\rm true}(r;q)$
exhibits much less scatter about unity as compared to the two previous
mass estimates. The asymptotic divergence at very large scales
($r\gtrsim 10$) is a numerical artefact of our crude deprojection algorithm.

\subsection{Comments and application to MS2137}\label{sec:discrepDISC}
\reffig{fig:mass-ratio} clearly shows that moderate values of the axis
ratio $q$ can lead to strong discrepancies between 2D and 3D mass
estimates or between lensing and X-rays or stellar kinematics.

It is difficult to fully characterize the ratio $M_{\rm lens}(r;q)/M_{\rm kin}(r;q)$
because it depends on the distribution of tracers $\nu(R,z)$ and is severely
sensitive to the orientation of the axis ratio relative to the line of sight.
Therefore a direct comparison between lensing and dynamical mass estimates
is hazardous. $M_{\rm lens}(r;q)/M_{\rm kin}(r;q)$ can have a different radial
behavior as a function of radius for a given axis ratio. It can be
either greater of less than unity.

Comparing lensing and X-rays mass estimates is easier since the X-rays mass
estimate is less sensitive to projection effects. In this respect
$M_{\rm lens}(r;q)/M_{\rm X}(r;q)$ will systematically be $>1$ (resp. $<1$)
for prolate (resp. oblate) halos with a well known radial behavior.

In the case of MS2137, a prolate halo with $q\sim0.4$ could well
explain most discrepancies between our best fit models and the results of
Sa04 and Al01. A prolate halo aligned toward the line of sight is a
natural explanation for the high concentration parameter we found
$c=11.73\pm0.55$ and may also explain the high concentrations  $c\approx 22$
in CL0024 \citep{kneib03} and $c=13.7\mypm{1.4}{1.1}$ in A1689
\citep{broadhurst05b}. Recently, \citet{oguri05} have investigated the effect
of triaxiality in A1689 and have similar conclusions as well as
\citet{clowe04} who studied numerical simulations
\citep[see also][]{piffaretti03}.

At this level, it is not possible to simply refine the modelling of
MS2137, since our prolate model is idealized. It should be triaxial
and/or not perfectly aligned with the line of sight just because the
projected density profile is elliptical. However the hypothesis of a
projected triaxial halo also provides a direct explanation for the 
misalignment between the projected diffuse stellar component of the cD
and the projected dark matter halo $\Delta \psi= 13.0\pm0.5$ deg.
\citet{binney85} and \citet{romanowsky98} give the necessary formalism to
infer the position angle and projected ellipticity of both dark and
luminous halos from their tridimensional triaxial shape and orientation.
The information that can be derived from the geometry of projected light
and dark matter densities is detailed in appendix
\ref{append:triax}. Basically, these independent constraints give the
following results for the orientation $\theta$ (polar angle of the
major axis with respect to the line of sight), the minor axis ratios
$c_{\rm DM}$ and $c_*$ of dark matter and stellar components
respectively: $\theta=27.4\pm5.1$, $c_{\rm DM}=0.55\pm0.08$,
and $c_*=0.52\pm0.12$. It is remarquable that this geometrical
information is fairly consistent with the value of $c_{\rm DM}\sim0.4$
and a perfect alignement ($\theta=0$) we assumed to explain the mass
discrepancies.

There is sufficient material to be convinced that no simple coupling
between 2D and 3D mass estimates is possible. Consequently, we expect
that most of the previous analyses based on such a coupling should be
considered with caution, either in terms of significance or in terms
of possibly biased results.

\section{Discrepancies : an expected general trend}\label{sec:discrepGEN}
The aim of this section is to predict the statistics of such mass discrepancies
between lensing and any other mass estimate which is not much sensitive to
asphericity effect like X-rays. Let us now consider a more general situation with
a triaxial halo $\rho(m)$ where $m^2=x^2+y^2/b^2+z^2/c^2$ with $0<b<c\le1$ and an
orientation relative to the line of sight parameterized by the polar angles
$(\theta,\phi)$ or the unit vector $\vec{n}$.

The mass within the sphere of radius $r$ is independent on the halo
orientation and reads 
\begin{equation}\label{eq:M3triDEF}
  M_{\rm true}(r;b,c) = \frac{1}{4\pi}\int_0^{2\pi} \der \varphi \int_0^\pi \sin \vartheta \der\vartheta\, \frac{M_3(\nu r;1,1)}{\nu^3}\,,
\end{equation}
with $\nu^2= \sin^2\vartheta \left( \sin^2\varphi\, /b^2 + \cos^2\varphi\,/c^2\right)+ \cos^2\vartheta$ and again $M_{\rm true}(R;1,1)$ is simply
the mass within radius $r$ for a spherically symmetric halo.

The mass $M_2$ within cylinder of radius $r$ will depend on their axis
ratios and the orientation $\vec{n}$ but the system is equivalent to
an elliptical projected mass distribution with axis ratio $\tilde{q}$ and position
angle $\psi$. Thus we can express $M_2$ as:
\begin{equation}\label{eq:M2triDEF}
  M_2(r;b,c,\vec{n}) = \frac{\tilde{q}_x^2}{\sqrt{f}} \int_0^{2\pi} \frac{\der \varphi}{2\pi} \frac{M_2(\tau r/\tilde{q}_x,1)}{\tau^2}\,
\end{equation}
where $\tau^2=\sin^2 \varphi + \cos^2 \varphi/ \tilde{q}^2$ and
$\tilde{q}$, $\tilde{q}_x$ and $f$ are given by Eqs. \eqref{eq:projQpsi}
in appendix \ref{append:triax}. They depend on the intrinsic axis ratios
and orientation. $M_2(r;1,1)$ is simply the cylindric mass within radius
$r$ for a spherically symmetric halo.

As before, an observer measuring the 3D mass profile within radius $r$
will find a different normalization as compared to an observer interested
in the cylindric mass of radius $r$. They will differ by a factor
\begin{equation}
\eta(R;b,c,\vec{n})=\frac{M_2(r;b,c,\vec{n})/M_2(r;1,1)}{M_{\rm true}(r;b,c)/M_{\rm true}(r;1,1)}\,.
\end{equation}
We can now calculate the statistical properties of this ratio by averaging
over the $b$ and $c$ PDFs of \citet{jing02tri} given by \eqref{eq:AXpdfJing}
and the orientation of the major axis (assumed isotropic). This can be
expressed as :
\begin{multline}\label{eq:etastat}
  p_R(\eta)=\frac{1}{4\pi}\int_0^1\der c\, p(c) \int_0^1 \der b\, p(b\vert c) \\\int_0^{\pi} \sin \theta \der \theta\, \int_0^{2\pi}\der \phi\, \delta\left[\eta-\eta(R;c,b,\vec{n})\right]\;.
\end{multline}

\begin{figure}[htbp]
  \resizebox{\hsize}{!}{\includegraphics{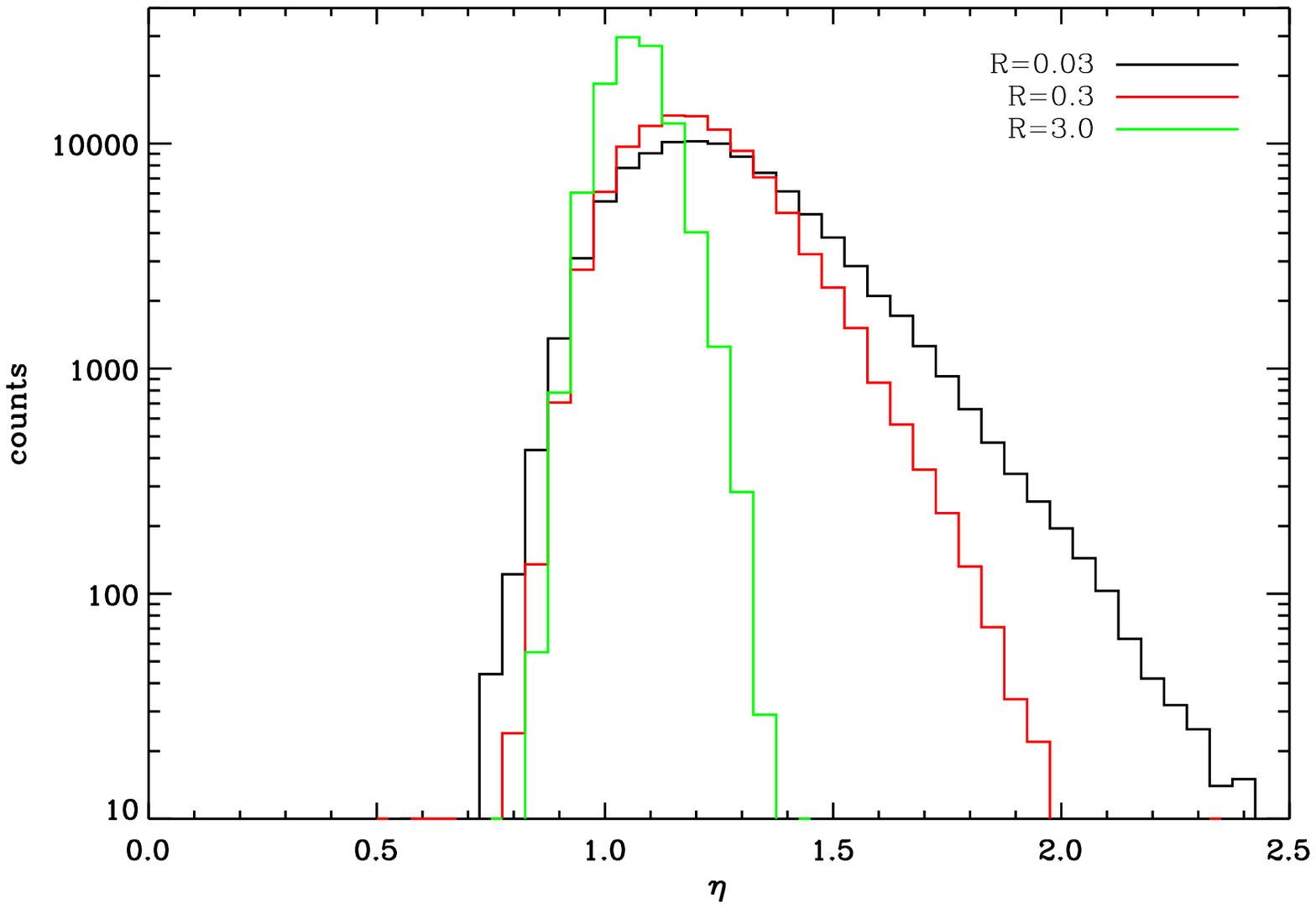}}
  \resizebox{\hsize}{!}{\includegraphics{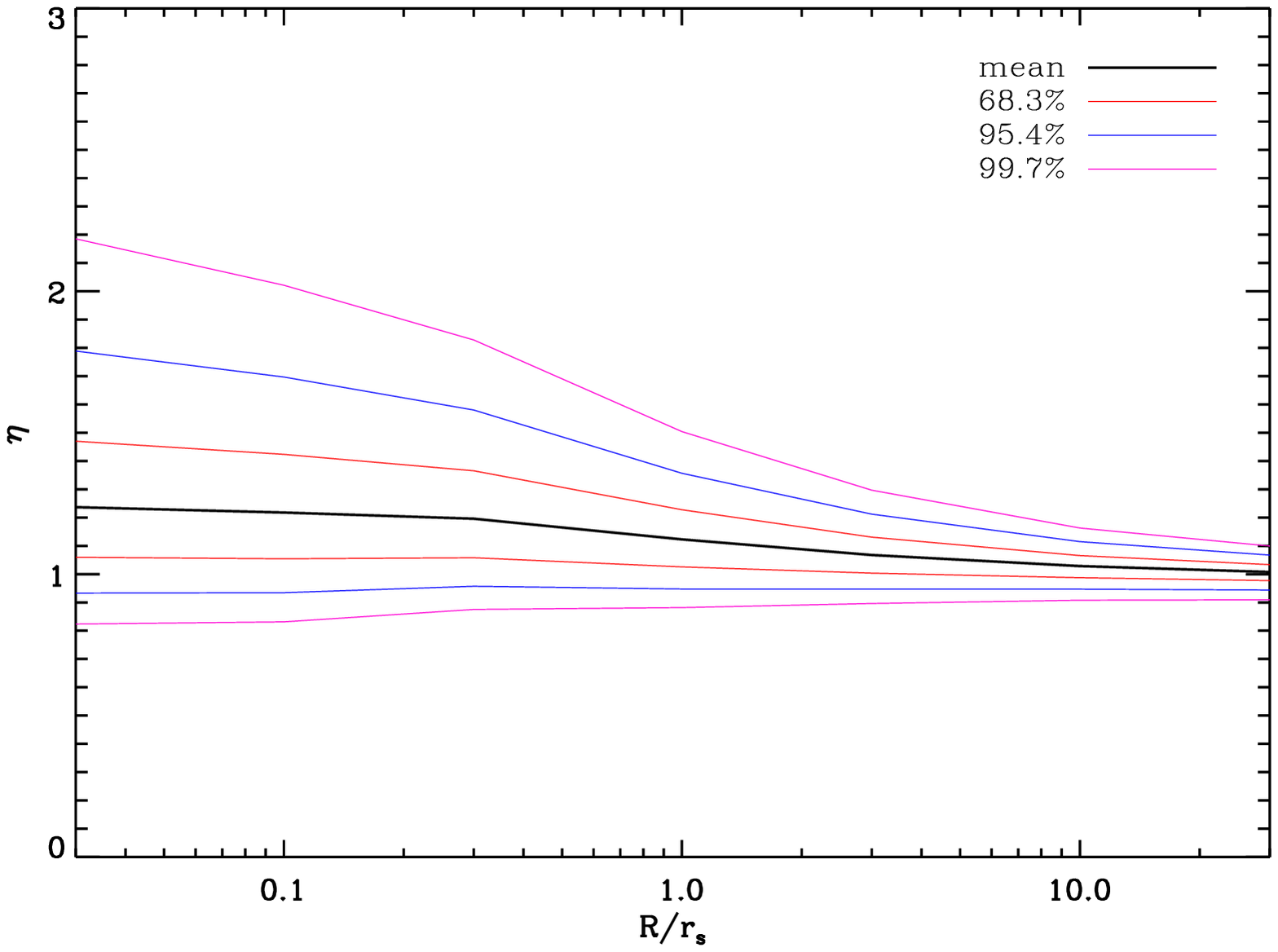}}
  \caption{\scriptsize {\it Upper panel : } Distribution for the mass
    ratio $\eta$ for three different radii $r=0.03$ (the broader black
    curve), $r=0.3 $ (the intermediate red curve) and $r=3.0$
    (the narrower green curve). At small scales, the distribution
    is broad and clearly not centered on $\eta=1$, leading to
    unreliable direct normalization between 2D and 3D mass
    estimates like in strong lensing and X-rays/stellar dynamics
    comparisons. With increasing radius ($r\gtrsim3.0$), departs
    significantly vanish and explain the overall agreement between
    large scale weak lensing and X-rays mass estimates. {\it \ \ Lower
      panel : } Mean (thick black curve) and 68.3, 95.4 and 99.7 \%
    quantiles (thin red, blue and magenta curves respectively)
    of the $\eta$ distribution as a function of radius. 
  }
  \label{fig:eta-dist}
\end{figure}
We plot on the upper panel of \reffig{fig:eta-dist} the distribution of $\eta$
for three fiducial values of $R=0.03,0.3$ and $3$ which are relevant
for strong lensing/stellar dynamics, strong lensing/X-rays and
weak lensing/X-rays comparisons respectively. We clearly see a broad,
shifted and skewed distribution which converges toward unity with increasing
radius. However, at small scales, the median value of $\eta$ is not unity
and readily extends toward high values $\eta\gtrsim1.5$. Typically
$\eta=1.24\mypm{0.23}{0.18}$ (resp. $1.19\mypm{0.17}{0.14}$, $1.07\pm0.06$)
for $R=0.03$ (resp. $0.3$, $3.0$). Thus, important departs between $M_2$
and $M_{\rm true}$ are naturally expected if halos are effectively triaxial.

\begin{figure}[htbp]
  \resizebox{\hsize}{!}{\includegraphics{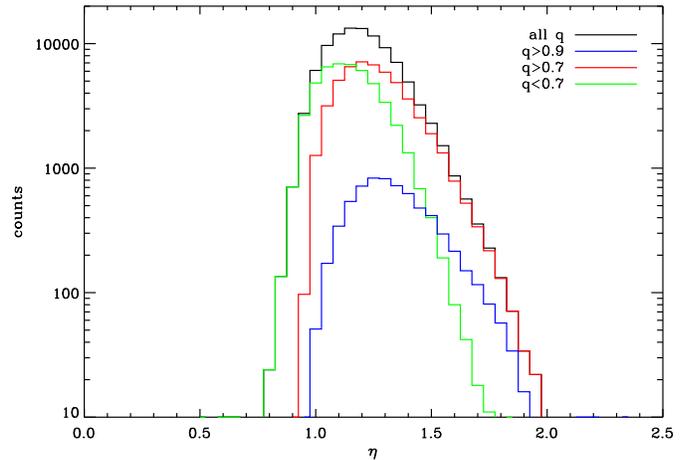}}
  \caption{\scriptsize For $r=0.3$, we show the dependence of $\eta$
    on the projected axis ratio $q$. The conditional PDFs $p(\eta\vert
    q>0.9)$ (blue) $p(\eta\vert q>0.7)$ (red) and $p(\eta\vert q<0.7)$
    (green) are compared to the overall distribution $p(\eta)$
    (black). The rounder the halo is, the more likely is the major
    axis of the halo oriented toward the line of sight and the higher is $\eta$.
  }
  \label{fig:eta-dist-cond}
\end{figure}

Moreover, there must be a correlation between the observed projected
axis ratio $q$ and $\eta$ since the apparently rounder halos are
likely to be elongated along the line of sight. This effect can be
seen on \reffig{fig:eta-dist-cond} where we plot the conditional PDFs
$p(\eta)$, $p(\eta\vert q>0.9)$, $p(\eta\vert q>0.7)$ and
 $p(\eta\vert q<0.7)$ for a radius $R=0.3$. The highest values
of $\eta$ are due to the roundest projected halos.
For instance, given $q>0.7$ we have $\eta=1.25\mypm{0.17}{0.13}$.

Finally projection effects of triaxial halos have the interesting properties
to explain the fact that weak lensing and X-rays measurements generally match 
since $\eta\sim1$ for $r \gtrsim$ a few $r_s$. Likewise the general
trend for strong lensing mass estimates to generally be greater
(by a factor of $1-3$) than X-rays since they occur at scales
$0.1 \lesssim r \lesssim1$ \citep{allen01,wu00}. The relative
normalization between lensing and stellar kinematics is more complex
and cannot be representated by the $\eta$ statistic. However we expect
a similar scatter and a strong dependence on the major axis orientation.

\section{Discussion \& Conclusion}\label{sec:discussion}
Regarding the particular case of MS2137, using a detailed modelling of
both strong and weak lensing data, we have shown that the dark matter 
density profile must be close to NFW. See also
\citep{miralda95,bartelmann96,gavazzi03,bartelmann03b,dalal03}
for similar conclusions. We have explained the reason why the Sa04 lens
model is inconsistent with lensing data (radial arc counter image,
weak lensing...). We have highlighted a possible discrepancy
between our lens model and other mass estimates from stellar kinematics
in the central cD galaxy and X-rays.

We have undertaken a thorough dynamical analysis of the line-of-sight
velocity distribution of stars in the cD in order to check whether or
not departs from gaussianity may explain the relative inconsistency
between our lens models and stellar kinematics. The effect of
non-gaussian aborption lines is to slightly lower ($\sim 15\%$)
the measured velocity dispersion estimates of Sa04 but does not greatly
improve the agreement between our lens model and the bias-corrected data.
Moreover such a bias cannot explain the disagreement between our lens
model and Al01 X-rays mass estimates. This latter 3D mass estimates turns
out to be consistent with stellar dynamics, showing that there must exist
some problem in the relative normalization of 2D and 3D mass estimates. 

These discrepancies can be alleviated if one considers the possibility
of departs from spherical symmetry for the dark matter and stellar
components. More precisely, we have shown that a prolate halo with its
major axis oriented close to the line of sight and an axis ratio $\sim0.4$
is likely to explain the discrepancies. This hypothesis is supported by
the misalignement ($\Delta \psi \sim 13^\circ$) between projected DM and 
stellar distributions.

Furthermore such a geometrical configuration well explains the concentration
parameter we infer from lensing analysis $c=11.73\pm0.55$. A prefered
elongated halo toward the LOS boosts lensing efficiency
\citep{bartelmann95b,oguri03,clowe04} and may explain the high
concentration of some strong lensing clusters
\citep{kneib03,broadhurst05b,oguri05,hennawi05b}. 

We have shown that triaxiality is a general problem that hampers any
attempt to simply couple 2D and 3D mass estimates assuming spherical
symmetry. Once projected, triaxial halos are elliptical and lens modelling
is able to take ellipticity into account. Usually dynamical or X-rays
analyses do not fully incorporate such a complexity. This should be an
important work to do before comparison to (or coupling with) lensing.
In Sect. \ref{sec:discrepGEN}, we have assumed the statistical distribution
of axis ratios proposed by \citet{jing02tri} in order to calculate the mass
$M_2(r)$ within cylinder of radius $r$ and the mass $M_{\rm true}(r)$ within
the sphere of same radius. The difference is important and can lead to a
significant discrepancy in the relative normalization between 2D and 3D
mass estimates.

The statistic of $\eta(r)$ shows that, at small scales $R<1$, in average
a systematic depart from unity is expected for $\eta$ with an important
scatter and skewness toward high values of $\eta$. Therefore the relative
normalization at small scales is biased and highly uncertain if one neglects
projection effects. At larger scales, the distribution of $\eta$ converges
to unity and explains why weak lensing mass estimates are generally in better
agreement with X-rays or dynamics of galaxies in clusters 
\citep[{\it e.g.}~][]{allen98,wu00,arabadjis04}. Similarly, the coupling
between stellar kinematics and strong lensing at clusters scales
\citep{sand02,sand04} or at galaxies scales
\citep[{\it e.g.}~][]{koopmans02,treu04,rusin03} may be oversimplistic
since they do not take asphericity into account. First, the mean value
$\eta \sim 1.2$ for $r\lesssim 0.3 r_s$ leads to a expected systematic bias,
but also the $\sim 20\%$ scatter in the distribution of $\eta$ will increase
the uncertainty in the mass normalization and prevent the appealing
temptation to couple these independent mass estimates.

\ \\
In conclusion, the density profile of the dark matter halo of MS2137-23
is well consistent with NFW and previous claimed discrepancies
may be due to the spherical symmetry assumption. Indeed, it turns out that
when coupling lensing to other mass estimates we cannot avoid a detailed
(and cumbersome) 3D triaxial modelling of X-rays and dynamical properties.
It is worth noticing that such a level of refinement is already achieved
in lensing studies that assume elliptical symmetry. The triaxiality of dark
matter halos (and stellar components) is a major concern for joint modelling
and should systematically be taken into account for future analyses.
As well, it is possible that X-rays or optically selected clusters are
biased toward elongated configurations, leading to an overefficiency for
lensing. The increasing precision of observations makes the assumption of
spherical symmetry abusive. Since clusters of galaxies are often seen as
an important cosmological probe. It is important to better characterize
their properties (mass, temperature, shape, abundance...) with realistic
triaxial symmetries.

\begin{acknowledgements}
I would like to acknowledge J. Miralda-Escud\'e who helped me starting
this work, which greatly benefited of his insightful advices.
I also thanks fruitful discussions with B. Fort, Y. Mellier and G. Mamon.
I am thankful to D. Sand who kindly made the velocity dispersion data
available and to I. Tereno for his help in the handling of MCMCs. Most of
this work has benefited of the TERAPIX computing facilities at IAP.
\end{acknowledgements}

\begin{scriptsize}
  \bibliographystyle{aa}
  \bibliography{references}
\end{scriptsize}

\appendix

\section{LOSVD of stars in the BCG}\label{append:dynam}
 The aim of this analysis is to derive the whole
line-of-sight velocity distribution (LOSVD) of stars from the
gravitational potential  $\Phi$. $r$ denotes the three-dimensional
radial coordinate whereas $R$ is the 2D projected radius and the density
of tracers $\rho_*(r)$ (the luminosity density).
We assume that the distribution function DF $f(\vec{r},\vec{v})$ can be
modeled by Osipkov-Merritt \citep{osipkov79,merritt85} distribution
functions (DF) which depend on the reduced energy $\rede=\Psi(r)-v^2/2$ and
angular momentum $L=rv \sin \zeta$ through the variable 
\begin{equation}\label{eq:qanis}
  Q=\rede -  \frac{L^2}{2 r_a^2}= \Psi(r) - \frac{v^2}{2}\left(1+
    \frac{r^2}{r_a^2}\sin^2\zeta\right)\;.
\end{equation}
In these equations $\Psi(r)=\Phi(r_{\rm max})-\Phi(r)$ is the reduced
potential, $\zeta$ is the polar angle of the velocity direction with
respect to $\vec{r}$ and $r_{\rm max}$ is the outermost radius at
which a particle is bound to the system, {\it i.e.} satisfying
$\rede \ge 0$. Otherwise specified, we set $r_{\rm max}=2\hmMpc$
in the following. $r_a$ is the anisotropy radius. Orbits are nearly
isotropic for $r_a\rightarrow \infty$ and nearly radial for $r>r_a$.

For Osipkov-Merritt models, the DF $f(Q)$ can directly be calculated,
through the Eddington formula \citep{binney87}
\begin{equation}\label{eq:eddingtonANIS}
  f(Q) = \frac{1}{\sqrt{8}\pi^2} \left[ \int_0^Q
    \frac{\der^2 \tilde{\rho}_*}{\der \Psi^2} \frac{\der \Psi}{\sqrt{Q-\Psi}} +
    \frac{1}{\sqrt{Q}} \left(\frac{\der \tilde{\rho}_*}{\der \Psi}\right)_{\Psi=0}\right]\;.
\end{equation}
where $\tilde{\rho}_* = (1+\frac{r^2}{r_a^2})\rho_*$.

Once the Eq. \eqref{eq:eddingtonANIS} numerically integrated, it is
possible to derive the LOSVD $p(R,v_\shortparallel)$
as a function of the projected radius $R$ by integrating over the line
of sight coordinate $z$ and over the perpendicular velocity $v_\perp$
with $v^2=v_\shortparallel^2+v_\perp^2$.
\begin{equation}\label{eq:velodist}
  p(R,v_\shortparallel) \propto \int_0^{z_{\rm m}} \der z
  \int_0^{v_{\perp,{\rm m}}} v_\perp \der v_\perp\: f(Q) \;,
\end{equation}
with $z_{\rm m}$ the maximum line-of-sight coordinate for a particle moving
at velocity $v$ located at the projected radius $R$ and satisfying
$\Psi(\sqrt{R^2+z_{\rm m}^2})=v^2/2$. In the isotropic case,
Eq. \eqref{eq:velodist} can be simplified :
\begin{equation}\label{eq:velodistISO}
  p_{\rm iso}(R,v_\shortparallel) = 2 \pi
  \int_{v_\shortparallel^2}^{2\Psi(R)} \der v^2 \int_0^{z_{\rm m}(v)}
  \der z\: f(\rede)\;,
\end{equation}
and a numerical integration is rather fast. However, in the general
case, this is not possible and we present in the following a much
faster Monte-Carlo technique.

The integration of Eq. \eqref{eq:velodist} is done by randomly
sampling the distribution function with a large number $N$ of stars.
Since the stellar density profile is known to a scaling mass-to-light
ratio, one can assign a radius $r$ to each star according to the
cumulative Hernquist stellar mass profile
$M_*(r)=M_*\left(\frac{r}{r+r_{s*}}\right)^2$. Each radius $r_i$ can
be projected onto the plane of sky yielding $\vec{R}_i$ and $z_i$,
the line of sight coordinate as before. At this point, it is trivial
to incorporate the smearing due to observational conditions like seeing
by adding a random displacement $\vec{R}_i \rightarrow \vec{R}_i + \vec{\delta
  R}_i$\footnote{where $\vec{\delta R}$ may follow a 2D Gaussian
  distribution with standard deviation $\sigma_{\rm
    seeing}=\mathrm{FWHM}/2.35$.}. Similarly, if the slit width
$\Delta$ is negligible $R_i=\vert\vec{R}_i\vert$ can be identified to
the position along the slit, otherwise, it is straightforward to split
$\vec{R}_i$ into $(x_i,y_i)$, only consider those points satisfying $2
\vert y_i\vert \le \Delta$ and then identify $x_i$ as the position
along the slit. This is the situation we shall consider in the following.

This spatial sampling of the DF is thus independent of the potential
$\Psi$ or the anisotropy radius $r_a$ and can be stored for
further calculation. For a given $\Psi(r)$ and $r_a$, one must solve
Eq. \eqref{eq:eddingtonANIS}, assign a velocity $v$ and a velocity
orientation $\Omega$ using the calculated DF $f(Q)$. This sampling is
done with acceptance-rejection technics \citep[{\it e.g.}~][]{press92}.
See also \citet{kuijken94} or \citet{kazantzidis04} for similar
applications. We can write the conditional PDFs for the
polar angle $\zeta$ and $Q$ at radius $r$ : 
\begin{equation}\label{eq:zetaPDF}
  \begin{split}
  p(\zeta \vert r) = &\frac{1}{2}\frac{\sin \zeta }{
    \left(1+\frac{r^2}{r_a^2}\sin^2\zeta\right)^{3/2}} \\
  p( Q \vert r) \propto & f(Q) \sqrt{2(\Psi(r)-Q)}\;.
  \end{split}
\end{equation}
Hence each star has a position $x_i$ and a line-of-sight velocity
$v_{\shortparallel,i}$. It is now possible to calculate
$p(R,v_\shortparallel)$ and the associated velocity dispersion
$\sigma_{\rm los}(R)=\sqrt{\overline{v_\shortparallel^2}}$ and kurtosis
$\kappa(R)=\frac{\overline{v_\shortparallel^4}}{\sigma_{\rm los}^4}-3$,
respectively related to the second and forth order moments of
$p(R,v_\shortparallel)$.

We now compute the LOSVD deduced from the best fit NFW model of MS2137 and
compare the inferred velocity dispersion to the measurements of Sa04. We assume
the same observational conditions {\it i.e.} a slit width
$\Delta=1.25\arcsec\simeq5.8\hmkpc$ and a gaussian seeing $0.6\arcsec=2.8\hmkpc$
FWHM. These data were obtained by assuming Gaussian absorption lines.
\citet{vandermarel93} showed that departs from Gaussianity
imply a bias in any velocity dispersion measurement. To the first order,
the biased pseudo-velocity dispersion $\vhalf$ reads:
\begin{equation}\label{eq:defBIAS}
  \vhalf = \frac{\slos}{1+\kappa/8}\;.
\end{equation}
In the Gaussian case ($\kappa=0$), $\vhalf$ reduces to $\slos$.

\begin{figure}[htbp]
  \resizebox{\hsize}{!}{\includegraphics{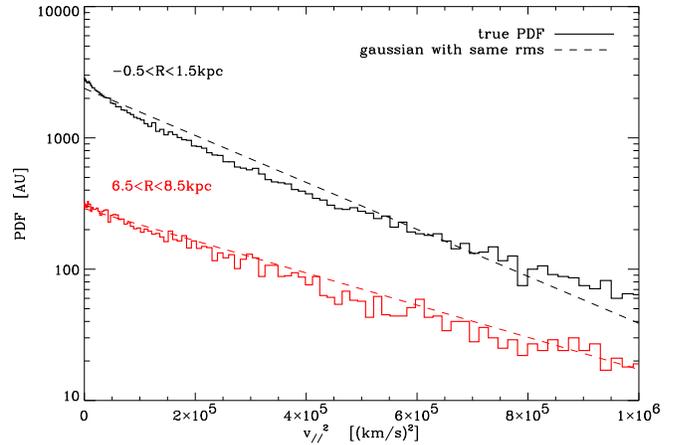}}
  \caption{\scriptsize
    Line-of-sight velocity probability distribution
    for stars with a projected radius in the innermost $-0.5<R<1.5\hmkpc$
    (the upper black histogram) and outermost bins $6.5<R<8.5\hmkpc$
    (the lower red histogram)of Sa04. In each case, we plot a gaussian
    distribution with the same dispersion for comparison. One can see
    non-gaussian tails for innermost stars. In this exemple,
    we consider the the best fit NFW model of Sect. \ref{sec:lensNFW}
    and orbits are isotropic.
  }
  \label{fig:velopdf}
\end{figure}

\reffig{fig:velopdf} shows the LOSVD as a function of the line-of-sight velocity
$v_\shortparallel$ for the innermost and outermost radial bins of Sa04. Departs from
Gaussianity are visible close to the center and decrease with increasing radius.
Therefore, the velocity bias changes with projected radius as can be seen
on the top panel of \reffig{fig:velopdf-compar}, in which we plot $\slos(R)$
and $\vhalf(R)$ for two values of the anisotropy radius $r_a=\infty$
and $r_a=10\hmkpc$.

\section{Further evidence for triaxiality}\label{append:triax}
In this section we follow the formalism of \citet{binney85} and
\citet{romanowsky98} and calculate the orientation and axis ratio of
a projected triaxial distribution as a function of its
intrinsinc 3D axis ratios $0\le c\le b\le 1$ such that the density
$\rho(\vec{r})=\rho(m)$ with $m^2=x^2+y^2/b^2+z^2/c^2$. We express the
orientation of the minor axis with the polar angles $(\theta,\phi)$
relative to the line of sight.
The projected density reads :
\begin{equation}\label{eq:proj2D}
  \Sigma(x,y)=\frac{2}{\sqrt{f}} \int_0^\infty \rho( u^2 + m^2)\,\der u\;,
\end{equation}
where 

\begin{gather}\label{eq:proj2D2}
  f=\sin^2 \theta \left( \cos^2 \phi+ \frac{\sin^2 \phi}{b^2}\right) + \frac{\cos^2\theta}{c^2}\;, \\\label{eq:proj2D2f}
  m^2 = \frac{1}{f}\left(Ax^2+B x y + C y^2\right)\;, \\
  A=\frac{\cos^2\theta}{c^2}\left(\sin^2\phi+\frac{\cos^2 \phi}{b^2}\right) + \frac{\sin^2\theta}{b^2}\;,\\
  B=\cos \theta \sin 2\phi\, \left(1-1/b^2\right)\frac{1}{c^2}\;,\\
  C=\left(\frac{\sin^2\phi}{b^2}+\cos^2 \phi \right) \frac{1}{c^2}\;.
\end{gather}
The projected distribution is elliptical with an axis ratio $\tilde{q}$
and a position angle $\psi$ given by :
\begin{subequations}\label{eq:projQpsi}
  \begin{align}
    \tilde{q}_{x/y}^2 = &\frac{2 f}{A+C \mp \sqrt{B^2+(A-C)^2}}\;,\label{eq:projQx}\\
    \tilde{q} = & \tilde{q}_y/\tilde{q}_x\;,\label{eq:projQ}\\
    \tan 2 \psi = & \frac{B}{A-C}\;.\label{eq:projPSI}
  \end{align}
\end{subequations}
These equations are verified by the dark matter and the stellar components 
which have their own axis ratios $c_{\rm DM}$, $b_{\rm DM}$, $c_*$ and $b_*$
but their principal axes are assumed to match. Generally, different values
of $c_i$ and $b_i$ lead to different values of $\tilde{q}_i$ and $\psi_i$.
This is what we observe in MS2137 where the light satisfies
$\tilde{q}_*=0.83\pm0.12$, $\psi_*=(18\pm1)\,\mathrm{deg}$ and our NFW
lens modelling yields $\tilde{q}_{\rm DM}=0.750\pm0.005$ and
$\psi_{\rm DM}=(4.90\pm0.15)$ deg. Hence we can infer the parameters
$\theta,\phi,c_{\rm DM},b_{\rm DM},c_*,b*$ from these constraints and
some additional priors since the problem is underconstrained. We can use
the axis ratio distribution found in cosmological simulations by
\citet{jing02tri}:
\begin{subequations}\label{eq:AXpdfJing}
  \begin{align}
    p(c) =& \frac{1}{\sqrt{2\pi}\sigma_c} \exp\left( -\frac{(c-\bar{c})^2}{2\sigma_c^2}\right)\;, \\
    p(b\vert c) = & \frac{3}{2 (1-\max(c,1/2))}\left[1-\left(\frac{2b-1-\max(c,1/2)}{1-\max(c,1/2)}\right)^2\right]\;,
  \end{align}
\end{subequations}
with $\sigma_c\sim0.113$ and $\bar{c}\sim0.54$. In addition, since the number
of constraints is not sufficient we force the intermediate axis ratios $b_*$
and $b_{\rm DM}$ to be equal to the most probable value. In other word, we have
\begin{equation}\label{eq:AXpdfMOD}
  p(b\vert c)= \delta\left(b-\frac{1+\max(1/2,c)}{2}\right)\;.
\end{equation}
The best fit with priors yields : $\theta=27.4\pm5.1$, $c_{\rm DM}=0.55\pm0.08$
and $c_*=0.52\pm0.12$. This analysis gives strong indications on the 
reliability of triaxial dark matter and stellar distribution with the major
axis relatively close to the line-of-sight and a value of $c_{\rm DM}$ close
to that inferred to explain the discrepancy between 2D and 3D mass estimates
in Sect. \ref{sec:discrepLENS}.

\end{document}